\documentclass[10pt,letterpaper]{article}
\usepackage[top=1.0in,left=0.9in]{geometry}

\usepackage{caption}

\usepackage{changepage}

\usepackage[utf8]{inputenc}

\usepackage{textcomp,marvosym}

\usepackage{fixltx2e}

\usepackage{amsmath,amssymb}

\usepackage{cite}

\usepackage{nameref,hyperref}

\usepackage[right]{lineno}

\usepackage{microtype}
\DisableLigatures[f]{encoding = *, family = * }

\usepackage{rotating}

\raggedright
\usepackage[aboveskip=1pt,labelfont=bf,labelsep=period,justification=raggedright,singlelinecheck=off]{caption}

\makeatletter
\renewcommand{\@biblabel}[1]{\quad#1.}
\makeatother

\date{}

\usepackage{lastpage,fancyhdr,graphicx}
\usepackage{epstopdf}
\fancyheadoffset[L]{2.25in}
\fancyfootoffset[L]{2.25in}

\begin{document}
\vspace*{0.35in}

\begin{flushleft}
{\Large
\textbf\newline{Intrinsic noise profoundly alters the dynamics and steady state of morphogen-controlled bistable genetic switches}
}
\newline
\\
Ruben Perez-Carrasco\textsuperscript{1*},
Pilar Guerrero\textsuperscript{1},
James Briscoe\textsuperscript{2},
Karen Page\textsuperscript{1}
\\
\bigskip
\bf{1} Department of Mathematics, University College London,\\ 
\quad Gower Street, London WC1E 6BT, UK.
\\
\bf{2} The Francis Crick Institute, Mill Hill Laboratories, London NW7 1AA, UK.
\\
\bigskip

* r.carrasco@ucl.ac.uk

\end{flushleft}
\section*{Abstract}
During tissue development, patterns of gene expression determine the spatial arrangement of cell types. In many cases, gradients of secreted signaling molecules -- morphogens -- guide this process. The continuous positional information provided by the gradient is converted into discrete cell types by the downstream transcriptional network that responds to the morphogen. A mechanism commonly used to implement a sharp transition between two adjacent cell fates is the genetic toggle switch, composed of cross-repressing transcriptional determinants. Previous analyses have emphasized the steady state output of these mechanisms. Here, we explore the dynamics of the toggle switch and use exact numerical simulations of the kinetic reactions, the corresponding Chemical Langevin Equation, and Minimum Action Path theory to establish a framework for studying the effect of gene expression noise on patterning time and boundary position. This provides insight into the time scale, gene expression trajectories and directionality of stochastic switching events between cell states. Taking gene expression noise into account predicts that the final boundary position of a morphogen-induced toggle switch, although robust to changes in the details of the noise, is distinct from that of the deterministic system. Moreover, stochastic switching introduces differences in patterning time along the morphogen gradient that result in a patterning wave propagating away from the morphogen source. The velocity of this wave is influenced by noise; the wave sharpens and slows as it advances and may never reach steady state in a biologically relevant time. This could explain experimentally observed dynamics of pattern formation. Together the analysis reveals the importance of dynamical transients for understanding morphogen-driven transcriptional networks and indicates that gene expression noise can qualitatively alter developmental patterning.

\section*{Author Summary}
Bistable switches are a common regulatory motif in biological processes. They consist of cross-repressing components that generate a switch-like transition between two possible states. In developing tissues, bistable switches, created by cross repressing transcriptional determinants, are often controlled by gradients of secreted signalling molecules - morphogens. These provide a mechanism to convert a morphogen gradient into stripes of gene expression that determine the arrangement of distinct cell types. Here we use mathematical models to analyse the temporal response of such a system. We find that the behaviour is highly dependent on the intrinsic fluctuations that result from the stochastic nature of gene expression. This noise has a marked effect on both patterning time and the location of the stripe boundary. One of the techniques, Minimum Action Path theory, identifies key features of the switch without computationally expensive calculations. The results reveal a noise driven switching wave that propels the stripe boundary away from the morphogen source to eventually settle, at steady state, further from the morphogen source than in the deterministic description. Together the analysis highlights the importance dynamics in patterning and demonstrates a set of mathematical tools for studying this problem.


\section*{Introduction}
Tissue development relies on the spatially and temporally organized allocation of cell identity, with each cell adopting an identity appropriate for its position within the tissue. In many cases, these cellular decisions are made by transcriptional networks controlled by extrinsic signals \cite{Lander2013, Green2015, Davidson2010}. These signals, usually termed morphogens, spread from a localized source within, or adjacent to, the developing tissue to form a spatial gradient that becomes the patterning axis of the tissue.  Cells are sensitive to the level of the morphogen and respond by producing a set of discrete gene expression stripes at different distances from the morphogen source \cite{Kicheva2012,Lander2013}. 

A transcriptional mechanism capable of the analogue to digital conversion necessary to transform the continuous morphogen gradient into distinct domains of gene expression is the so-called \emph{genetic toggle switch} \cite{Saka2007,Cotterell2010}. This network motif, present in many biological contexts, consists of cross-repression between sets of transcriptional determinants that are expressed mutually exclusively in alternative cell identities \cite{Enver2009 ,Zhou2011, Shu2013, Manu2009,Balaskas2012}. Thus the expression of one set of factors represses the alternative identity and vice versa, creating a bistable switch \cite{Cherry2000, Balazsi2011, Sokolowski2012}. This mechanism has been extensively studied as a way for cells to make decisions and produce distinct outputs in response to biological signals \cite{Wang2010, Lv2014, Guantes2008,Verd2014}. In the case of tissue patterning, a morphogen gradient can modulate the transcription rates of one or more genes that comprise the switch. This controls the position along the patterning axis at which the switch creates a boundary between cell identities \cite{Balaskas2012,Saka2007}. Moreover, the principle can be extended to incorporate multiple morphogen controlled toggle switches, each producing a boundary at a distinct position, hence explaining the multiple stripes of gene expression generated in a tissue \cite{Panovska2013,Tufcea2015}

Mathematical models of morphogen-controlled toggle switches that reproduce the ultrasensitivity necessary to create discrete gene expression boundaries also generate a temporal sequence of gene expression, prior to reaching steady state, that recapitulates the final spatial pattern \cite{Balaskas2012,Tufcea2015,Verd2014}. This sequence and its timing is a consequence of the inherent dynamical properties of the bistable switch \cite{Strogatz2014,Guantes2008,Tufcea2015}.  Strikingly, the temporal behaviour predicted by the models corresponds to experimental observations of gene expression timing in several developing tissues \cite{Balaskas2012,Tschopp2009,Francois2010,Jaeger2006,Zhang2012} and has led to the suggestion that it explains temporal features of morphogen controlled tissue patterning \cite{Tufcea2015}. 

Despite the apparent agreement between mathematical models and experimental observations, whether the models correctly identify the biological mechanism responsible for the dynamics and boundary positioning in morphogen-patterned tissues remains unclear. In particular, current models (excepting \cite{Zhang2012}) are deterministic and have not addressed whether stochastic fluctuations that arise from noisy gene expression qualitatively alter the behaviour of morphogen-controlled toggle switches. Fluctuations in the production and degradation rates of mRNA and protein molecules in individual cells can lead to substantial molecular heterogeneity \cite{Thattai2001,VanKampen2007,Kepler2001,Swain2002}.  Moreover, genes switch between active and inactive states resulting in bursts of transcription interspersed by refractory periods in which transcription is suppressed \cite{Blake2003,Raj2006,Pare2009,Thattai2001}. The intrinsic variations introduced by these processes could facilitate spontaneous transitions between different cell states, see for example \cite{Roma2005, Tian2006, Wang2007, Morelli2008, Mehta2008, Frigola2012, Lv2014, Tse2015}. Moreover, the effect of stochastic fluctuations on the position and precision of boundaries needs to be explored as previous work has suggested, counter-intuitively, that noise can sharpen boundaries in a tissue \cite{Lander2013,Zhang2012}. Thus in addition to the effect of gene expression noise on the steady state of a genetic toggle switch, understanding how stochastic fluctuations influence the temporal behaviour of switching along a morphogen patterning axis is necessary.  

Here we develop a theoretical framework to investigate the dynamics of morphogen-controlled genetic toggle switches and analyse how noise in gene expression affects tissue patterning by this mechanism. We show that exact numerical simulations of the kinetic reactions, the corresponding Chemical Langevin Equation, and Minimum Action Path theory provide insight into the trajectory, dynamics and directionality of stochastic switching between states of a bistable switch. The analysis predicts that intrinsic fluctuations diminish the influence of dynamical ghosts on the temporal evolution of the system so that they no longer dominate the time scale of pattern formation. Moreover, a new steady state position of the pattern boundary is predicted by the stochastic models, which is distinct from that suggested by deterministic analyses. This new boundary position is the consequence of a patterning wave that propagates along the pattering axis of the tissue and sharpens as it advances. The velocity and precision of this moving boundary are determined by intrinsic noise properties, controlled in the model by the typical number of proteins in the system and production burst sizes. Moreover, boundary propagation can be very slow and in some cases would not be expected to reach steady state in a biologically relevant time. Thus our analysis emphasizes the importance dynamical transients for understanding patterning processes and illustrates a theoretical toolkit to explore these questions.

\section*{Models}

\begin{figure}[ht!]
\centering
\includegraphics[width=0.75\textwidth]{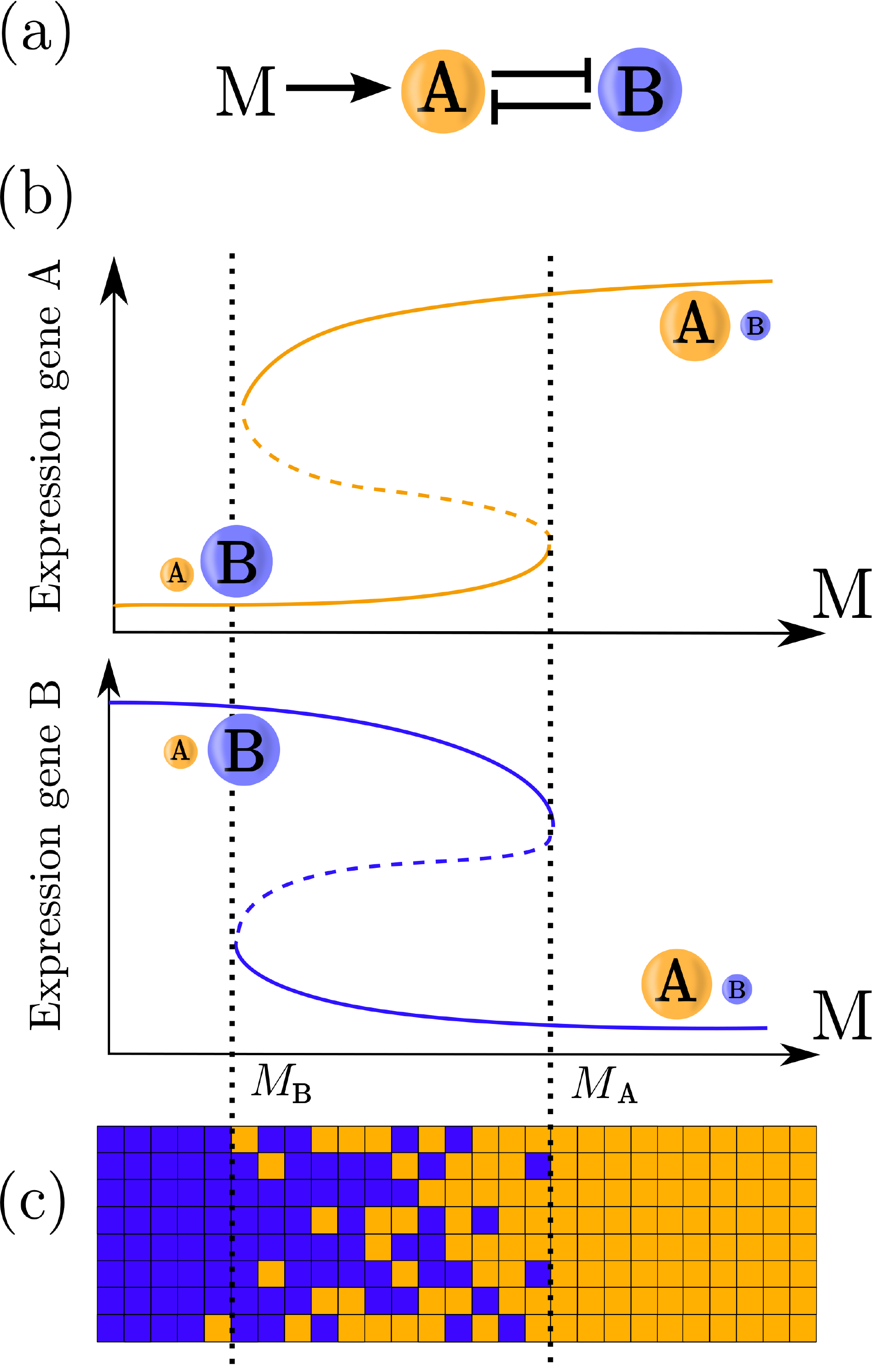}
\caption{ \label{fig.network} \textbf{Bistable switch patterning schematic.}  
 (a) Genetic network showing the cross-repression of genes $A$ and $B$. The signal $M$ activates gene $A$ changing the available steady states. (b)  Schematic stability diagrams for expression of genes A and B showing the available steady states for different values of the signal. Stable steady states are indicated with solid lines, while the unstable steady state (in this case a saddle point) is indicated with a dashed line. (c) Schematic of tissue patterning. The available states in the bistable zone will be determined by the history of the signal and by stochastic effects.}
\end{figure}

In order to characterise the dynamics of a bistable switch we consider a model in which two genes $A$ and $B$ repress each other, and a morphogen signal $M$ acts as an activator of gene $A$ promoting its expression (Fig. \ref{fig.network}). The two possible cellular steady state outcomes, A and B, are characterised by high levels of expression of one gene and lower levels of the other. For intermediate morphogen signalling levels, both states are stable, with the region of bistability given by $M_B<M<M_A$. In this region, the cellular outcome will depend on the initial expression of the cell and the history of the morphogen signal. In response to the graded distribution of the morphogen $M$, this model is capable of generating two abutting stripes of gene expression. This mechanism can be readily extended to accommodate additional stripes at different morphogen levels\cite{Balaskas2012,Tufcea2015,Panovska2013}. 

Even in the simple scenario, however, there are many possible mechanisms regulating two-gene interactions, these can act at transcription, translation or post-translational levels. Despite this, the essential mechanism of patterning reduces to the same bistable switch motif (Fig. \ref{fig.network})  \cite{Balaskas2012,Song2010,Tufcea2015,Zhang2012,Sokolowski2012,Cherry2000}.

In the current study we consider that protein production occurs on a much slower time scale than transcription factors binding and unbinding to the enhancer, RNA polymerase binding and unbinding the promoter, transcription and mRNA degradation. Under this assumption the production of each protein can be considered proportional to the probability of finding the polymerase bound to its promoter $\tilde p_i$ with $i\in{A,B}$. The rate of change in time of protein concentration $n_i(t)$ can then be described as

\begin{eqnarray}
\dot{n}_A&=&\alpha_A \tilde p_A(M,n_B)-\delta_A n_A\nonumber\\
\dot{n}_B&=&\alpha_B \tilde p_B(n_A) - \delta_B n_B, \label{eq.model}
\end{eqnarray}
where $\alpha_i$ is the protein production rate when RNAp is bound to gene $i$ , and $\delta_i$ is the effective degradation rate of protein $i$. Additionally, the polymerase binding probability $\tilde p_A$ depends on the morphogen signal M, since the morphogen controls the production rate of gene $A$. In Eq. \ref{eq.model} the concentration of protein is given in arbitrary units that can be related to the actual number of proteins $N_i$ through a multiplicative constant $\tilde\Omega$ \emph{i.e.} $N_i=n_i\tilde\Omega$.

Eq. (\ref{eq.model}) can be non-dimensionalised by expressing time in units of $\delta_A^{-1}$ and the concentration of proteins in units of $\alpha_A/\delta_A$ obtaining,

\begin{eqnarray}
\dot{x}_A&=&p_A(M,x_B)- x_A\nonumber\\
\dot{x}_B&=&\alpha p_B(x_A) - \delta x_B, \label{eq.modelnond}
\end{eqnarray}
describing the evolution of the non-dimensional protein expression $x_i\equiv \delta_A n_i/\alpha_A$ where the magnitudes $\alpha\equiv\alpha_B/\alpha_A$ and $\delta\equiv\delta_B/\delta_A$ are respectively the relative rates of production and degradation of genes $A$ and $B$. Similarly, the non-dimensional binding probabilities are defined as $p_A(M,x_B)\equiv\tilde p_A(M,n_B)$ and $p_B(x_A)\equiv\tilde p_B(n_A)$. This non-dimensionalization not only simplifies the study of the system but also reveals intrinsic properties of the network. For given functions $p_i$, the whole dynamical system behaviour only depends on the parameter ratios $\alpha$ and $\delta$ regardless of the actual values of $\alpha_i$ or $\delta_i$. Additionally, the non-dimensionless expression levels of each stable state ($\dot x_i(x_i^{st}) =0$) only depend on the ratio $\alpha/\delta$ independent of the actual values of $\alpha$ and $\delta$,
\begin{eqnarray}
x_A^{st}=p_A(M,x_B^{st})\nonumber\\
x_B^{st}=\frac{\alpha}{\delta}p_B(x_A^{st}). \label{eq.stst}
\end{eqnarray} 

\begin{figure}[ht!]
\centering
\includegraphics[width=0.78\columnwidth]{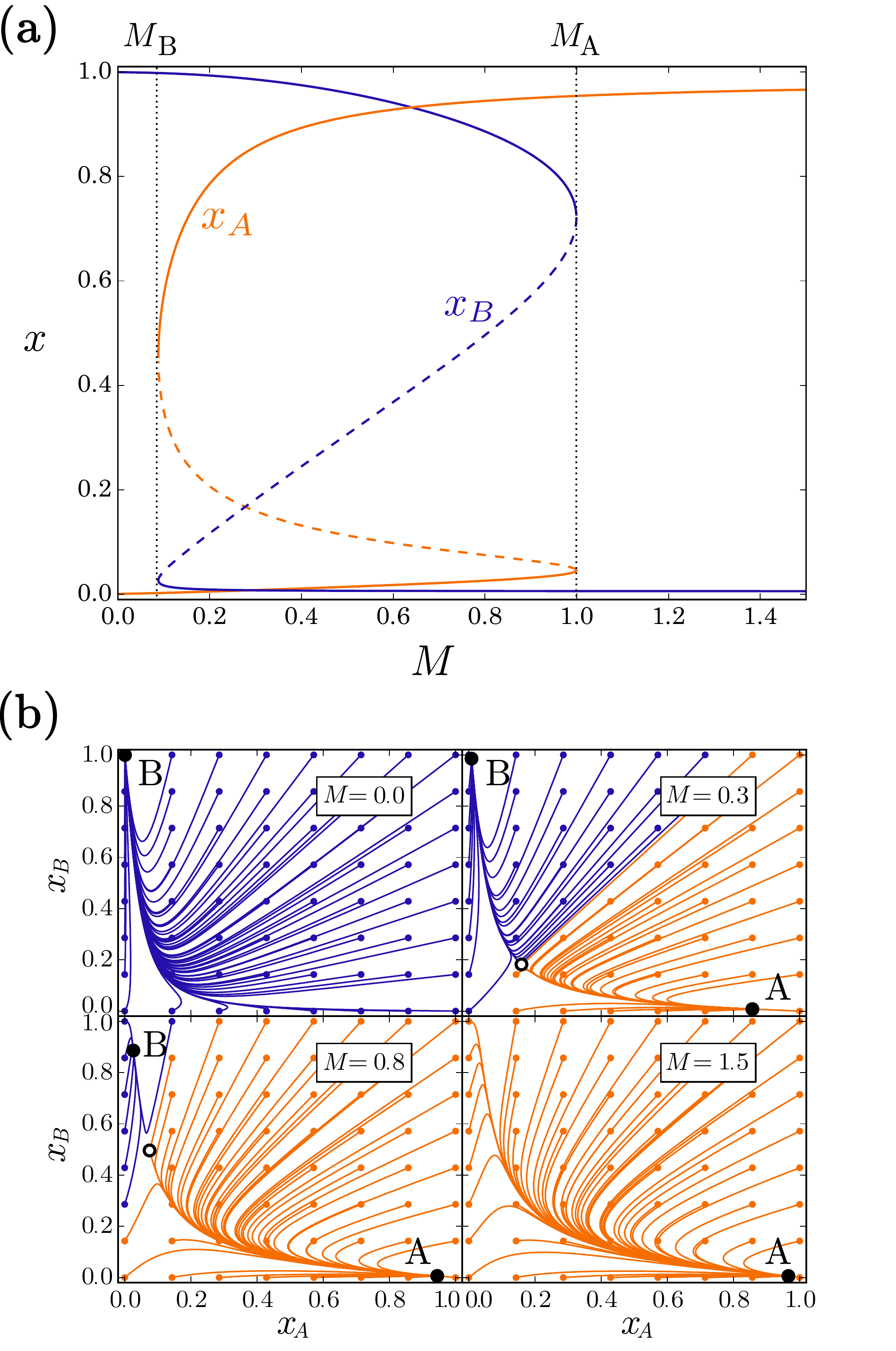}
\caption{ \label{fig.landscape} \textbf{ Dynamical behaviour of the bistable switch.} (a) Bifurcation diagram of the bistable switch, showing solutions of (\ref{eq.stst}) as a function of the signal $M$ for proteins $x_A$ (orange) and $x_B$ (blue). The stable steady states (solid lines) and the saddle points (dashed lines) are indicated. (b) Trajectories for different starting points marked by coloured circles. The colour indicates the final stationary state. Additionally, the stable steady states (black circles), and the saddle point (white circle) are indicated. Parameters used are $\alpha=\delta=\rho_A=1$,$\rho_B=1.75\cdot10^{-4}$, $K_A=10^{-3}$, $K_B=3\cdot10^{-2}$, $K_M=1$, $f=10.0$.}
\end{figure}

The RNAp binding probabilities $p_i$ can be described using statistical physics principles by computing the fraction of possible equilibrium configuration states in which RNAp is bound to the target gene promoter \cite{Sherman2012}. We will consider the case in which each promoter has two non-overlapping binding sites for the repressive transcription factor that interact independently and, when occupied, forbid the binding of RNAp \cite{Bintu2005}. Assuming that the signal reduces the recruiting binding energy of RNAp to the promoter of gene A, and that its effector also has two independent binding sites, also independent of the repressor sites, the promoter state can be written as,

\begin{eqnarray}
p_A(M,x_B)&=&\left(1+\rho_A\left(\frac{1+M/K_M}{1+fM/K_M}\right)^2\left(1+\frac{x_B}{K_B}\right)^2\right)^{-1}\nonumber\\
p_B(x_A)&=&\left({1+\rho_B\left(1+\frac{x_A}{K_A}\right)^{2}}\right)^{-1},\label{eq.reg}
\end{eqnarray}
where $\rho_i$ sets the basal gene activation, $f$ controls the activation strength of the signal, $K_i$ is the re-scaled equilibrium dissociation constant of protein $i$ for its binding site in the relevant enhancer \cite{Diploidy}, and $K_M$ the equilibrium dissociation constant of the morphogen effector with the enhancer of gene $A$. All parameters of (\ref{eq.reg}) are positive and $f>1$. The regulatory equations \eqref{eq.reg} return the usual polynomial ratio that gives rise to a sigmoidal response for the activator and the repressor \cite{Bintu2005,Guantes2008,Weiss1997} (Fig. \ref{fig.landscape}).

The morphogen gradient signal $M$ is typically a monotonically decreasing function of the distance from its source. Thus, in contrast to many studies of toggle switches, which focus on the behaviour for some fixed values of the signal, the patterning problem requires us to understand how the response varies along a continuous gradient of a signal. Without any loss of generality, the precise spatial dependence will be omitted and we just consider the outcome in response to a continuous signal $M$. Some previous studies introduce a more complex range of interaction between morphogen gradients and cells, either by considering spatial coupling between cells \cite{Zhang2012,Sokolowski2012} or by direct interpretation of the linear spatio-temporal changes of the morphogen gradient \cite{Richards2015}. In contrast, our choice to remove cell-cell interaction or signalling dynamics aims to reveal fundamental patterning properties of a bistable switch ubiquitous in similar multistable regulatory systems.

\subsection*{Stochastic dynamics}

Protein production and degradation described in Eq. (\ref{eq.model}) give a deterministic description of the bistable switch dynamics. This deterministic description is the coarse grained outcome of the underlying single production and degradation stochastic events that generate intrinsic noise in the network \cite{Thattai2001,VanKampen2007}. This random component can be included in the expression dynamics by the Chemical Langevin Equation (CLE) approximation, that introduces the intrinsic fluctuations as the addition of a multiplicative noise term to the deterministic description \cite{Gillespie2000},

\begin{eqnarray}
\dot{x}_A&=&p_A- x_A+\sqrt{\nu_A p_A+x_A}\,\xi_A(t\nonumber)\\
\dot{x}_B&=&\alpha p_B-\delta x_B+\sqrt{\nu_B\alpha p_B+\delta x_B}\,\xi_B(t) \label{eq.CLE}
\end{eqnarray}
where $\xi_i(t)$ is a Gaussian white noise with zero average and is delta-correlated:
\begin{eqnarray}
\langle \xi_i(t) \rangle &=& 0\nonumber,\\
\langle \xi_i(t)\xi_j(t')\rangle &=&\delta_A\alpha_A^{-1}\tilde\Omega^{-1}\delta(t-t')\delta_{ij}\nonumber\\
&\equiv&\Omega^{-1} \delta(t-t')\delta_{ij}.
\end{eqnarray}
Here $\delta_{ij}$ is Kronecker's delta, $\delta(t-t')$ is Dirac's delta, and $\Omega$ is the volume parameter relating concentrations with number of molecules ($N_A=n_A\tilde\Omega=x_A\Omega$), which also allows us to express the rates in terms of absolute changes in protein number per unit of time for the different reaction channels. The deterministic result (\ref{eq.modelnond}) is recovered in the limit $\Omega\rightarrow\infty$ \cite{Gillespie2000,VanKampen2007}. On the other hand, the parameters $\nu_i$ introduce the stoichiometries of the production reactions, accounting for the effects of expression bursts \cite{Gillespie2000}. 
  
The introduction of parameters $\nu_i$ and $\Omega$ does not perturb the deterministic landscape, which only depends on $p_i$, $\alpha$ and $\delta$. Different $\nu_i$ and $\Omega$ will result in different noise dependence with the regulatory functions and provide a natural mechanism to change fate determination while keeping the same macroscopic description (\ref{eq.model}).

The CLE (\ref{eq.CLE}) is the limiting description of the actual set of production and degradation reactions governing the change in protein numbers,

\begin{eqnarray}
\varnothing \xrightarrow{\tilde\Omega\alpha_A p_A/\nu_A} \nu_A A,&\quad&A \xrightarrow{\delta_A} \varnothing \nonumber\\
\varnothing\xrightarrow{\tilde\Omega\alpha_B p_B/\nu_B} \nu_B B,&\quad& B \xrightarrow{\delta_B} \varnothing. \label{eq.reaction}
\end{eqnarray}
The CLE approximation breaks down if the reaction rates change significantly within the time scale of the slowest reaction, and hence full simulation of the reaction kinetics (\ref{eq.reaction}) is required  \cite{Gillespie2000}. Here we will compare full simulations of (\ref{eq.reaction}) with the solution of the CLE to test the validity of the approximation for the biological scenario studied (for details on the implementation of CLE and kinetic reactions see \nameref{S1_Text}). 

The sources of noise described reproduce the effects of intrinsic fluctuations arising from production and degradation events of the proteins, through the stochastic parameters $\Omega$, $\nu_A$ and $\nu_B$. Other sources of noises, such as noise in the morphogen signal, may be of relevance in specific biological scenarios and will require specific formulations of kinetic equations and CLE dynamics. 

\section*{Results}
\subsection*{The patterning time of a deterministic bistable switch varies with morphogen level}

\begin{figure}[t!]
\centering
\includegraphics[width=0.9\columnwidth]{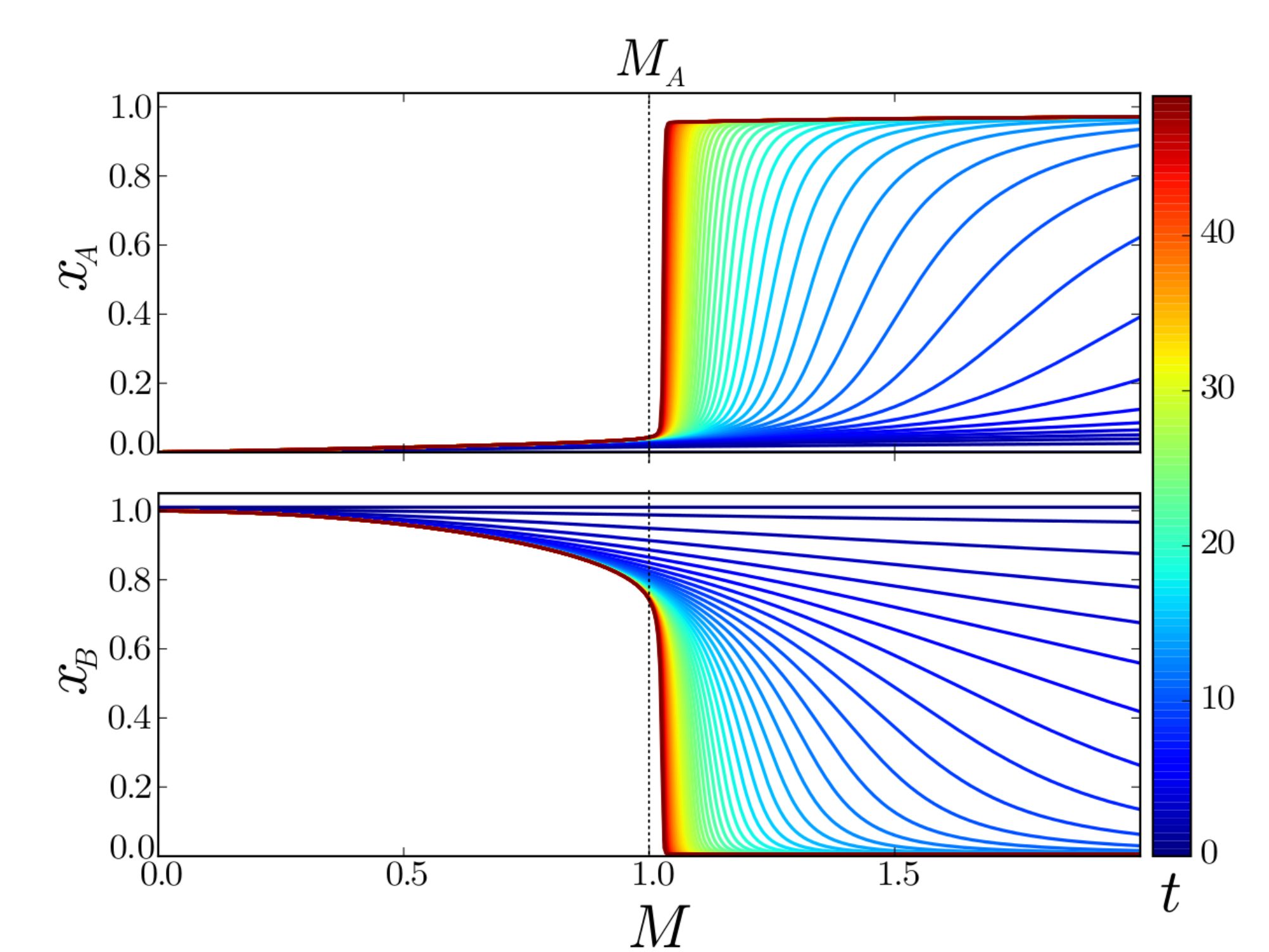}
\caption{\label{fig.xprofile} \textbf{Deterministic spatial pattern formation over time}. Initial conditions are $x_A=0$ and $x_B=1.0$. The rest of the parameters are those of Fig. \ref{fig.landscape}.}
\end{figure}

For a genetic toggle switch such as that described by Eq. (\ref{eq.model}), bistability means that the final steady state is determined not only by the level of the signal applied but also the initial condition of the system. In the absence of noise, the response of the system to the signal $M$ can be divided into a bistable regime ($M_B<M<M_A$) where the steady state is dependent on the initial conditions, and two monostable regimes ($M>M_A$ and $M<M_B$) where the final state of the system, $A$ or $B$, is independent of the initial conditions. The way in which the system switches steady state is by the application of a signal that is outside of the bistability zone, $M>M_A$ for an initial state B, and $M<M_B$ for initial state A (Fig. \ref{fig.landscape}). To conform to the tissue patterning paradigm, we assume that prior to the application of the morphogen signal, the ``constitutive" gene, $B$, is active homogeneously throughout the tissue. The establishment of the morphogen gradient results in the activation of the ``target" gene $A$ at positions at which the morphogen concentration exceeds the threshold ($M>M_A$), thereby creating the pattern (Fig. \ref{fig.xprofile}). 

\begin{figure}[ht!]
\centering
\includegraphics[width=0.9\columnwidth]{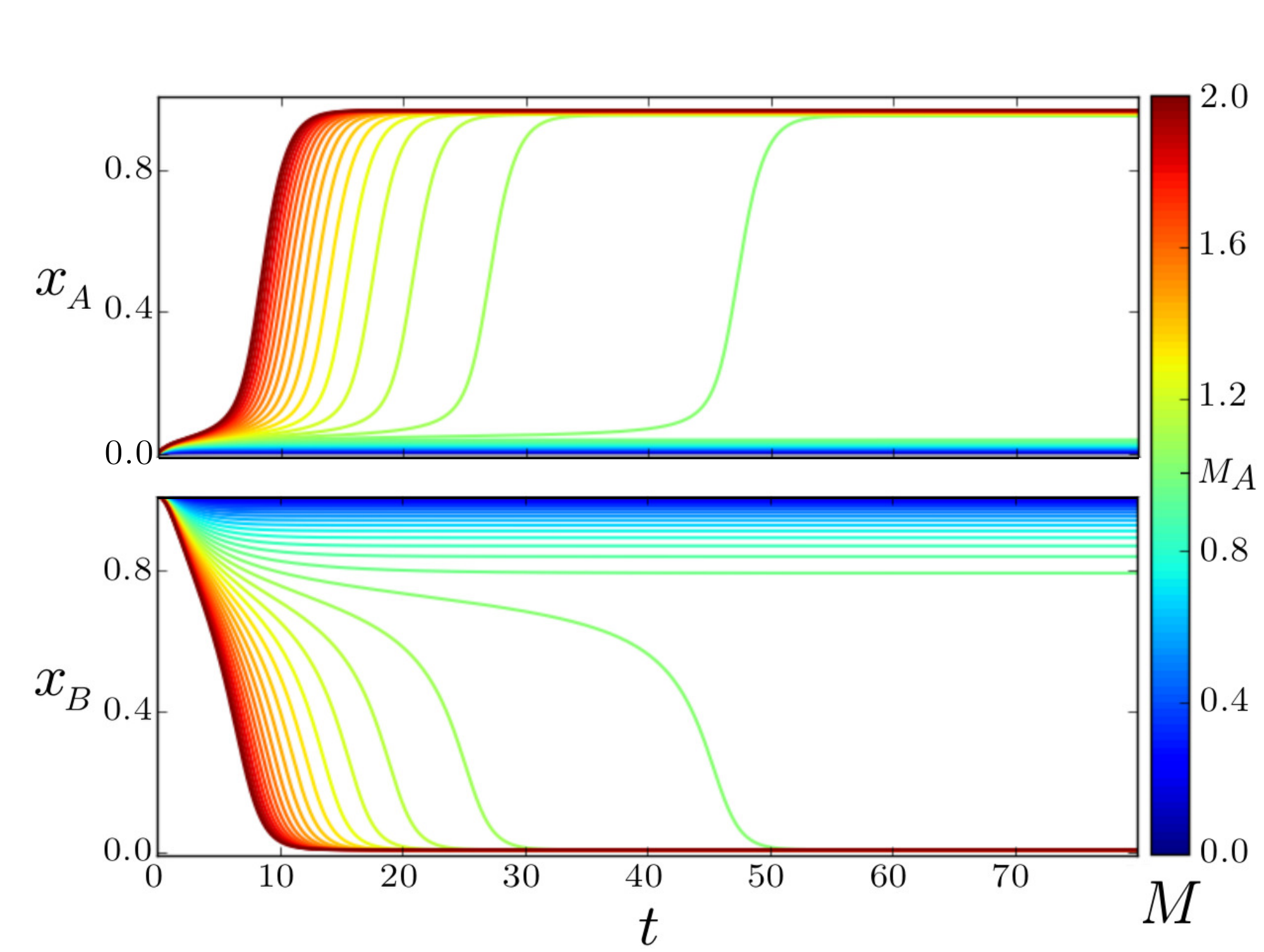}
\caption{ \label{fig.transientprofile} \textbf{Deterministic transient slows down close to the threshold $M_A$}. Transient expression profiles over time depend on the morphogen signal values. Initial conditions are $x_A=0$ and $x_B=1.0$. The rest of the parameters are those of Fig. \ref{fig.landscape}.}
\end{figure}

\begin{figure}[ht!]
\centering
\includegraphics[width=0.8\columnwidth]{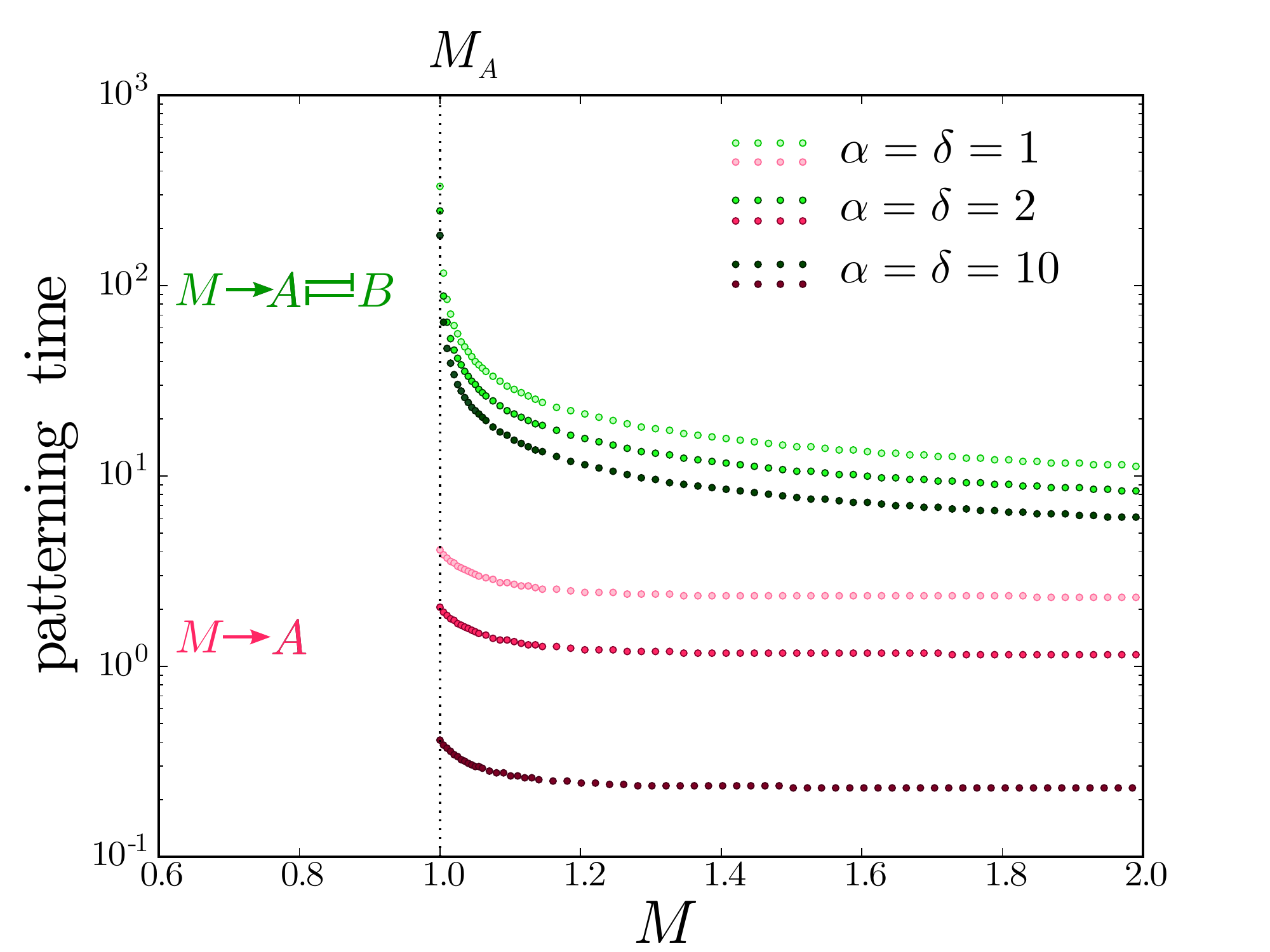}
\caption{\label{fig.patterningtime} \textbf{Variation of deterministic patterning time along the tissue is intrinsic to the bistable switch.}  The bistable switch (green circles) is compared with the non-feedback case (magenta circles) for three different values of $\alpha_B=\delta_B=\{1,2,10\}$. For signal values below the threshold signal $M_A$ (dotted line), state A is never reached. The patterning time is measured as the time necessary to observe a response $x_A>0.9$. Since $\delta_A$ is kept constant, the dimensional patterning time is not affected for different values of $\alpha$ and $\delta$.  Other parameters are the same as in Fig. \ref{fig.landscape}.}
\end{figure}

\begin{figure}[ht!]
\centering
\includegraphics[width=0.9\columnwidth]{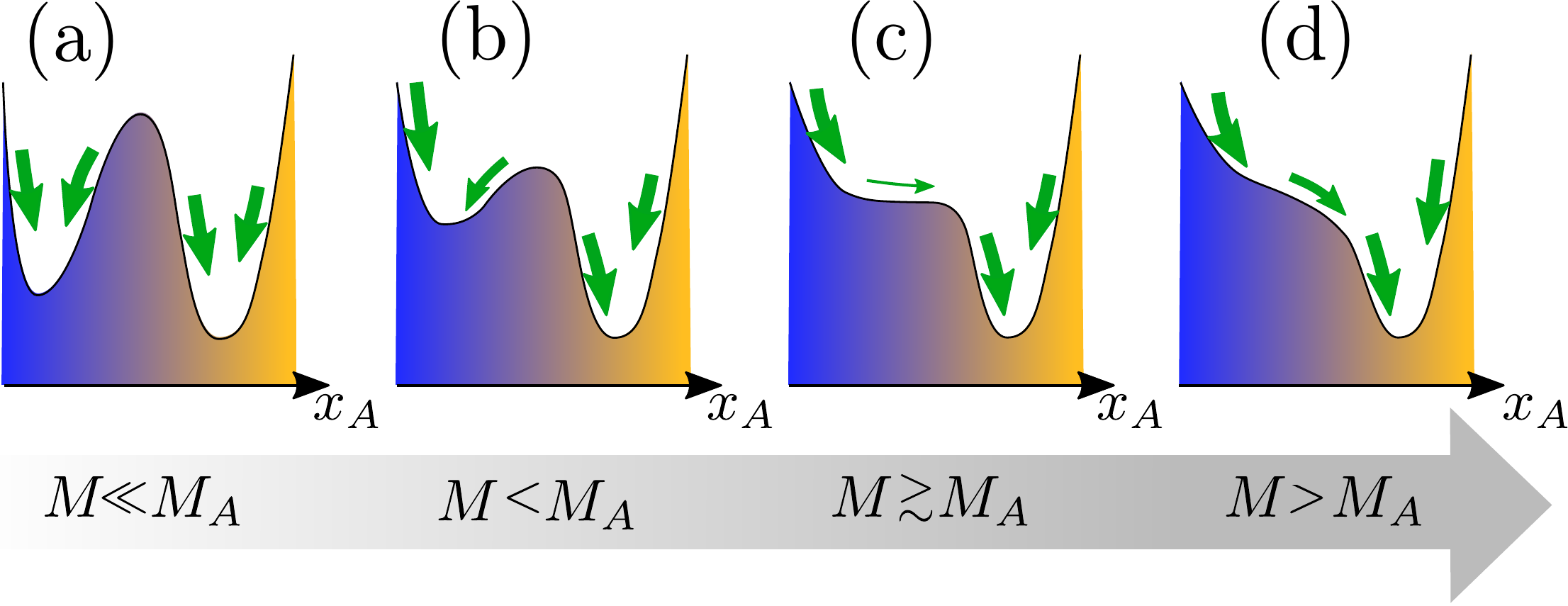}
\caption{ \label{fig.saddlenode} \textbf{ Schematic representation of the cell fate dynamics for different values of the morphogen signal $M$ around the threshold $M_A$}. For each morphogen value $M$, the velocity of change in genetic expression (green arrows) depends on the expression levels. (a) For low values of the signal in the bistable zone ($M_s\ll M\ll M_A$), there are two well defined cellular states. (b) As the signal increases, the attraction towards the stable state B becomes weaker. (c) At the threshold $M_A$ the stable minimum and the saddle collide cancelling each other (saddle-node bifurcation) resulting in a flat dynamical landscape for values $M\gtrsim M_A$ with a very slow change in gene expression in time. (d) For higher morphogen signal the evolution towards the activated state A becomes faster.}
 \end{figure}

The deterministic description of the patterning process is thus reduced to understanding the evolution towards the stable state A in zones of the tissue where $M>M_A$. For those positions, state A is not reached immediately but will undergo a transient expression described by Eq.(\ref{eq.modelnond}). We refer to the time it takes for the system at a certain signal $M$ to approach steady state as the patterning time, $T$. The higher the morphogen concentration the faster gene expression changes, saturating at a minimal characteristic patterning time $T_c$. In contrast, the patterning process becomes slower the closer the signal is to the threshold signal $M_A$, where levels of gene $A$ remain low for a long time before it is expressed ($T\gg T_c$) (Figs. \ref{fig.transientprofile} and \ref{fig.patterningtime}). This marked increase in the patterning time is a signature of the mechanism by which the stability of state B is lost at $M=M_A$ (saddle-node bifurcation); around $M_A$ the stability of $B$ varies smoothly with $M$ and hence close to $x^{st}(M_A)$ the dynamics are very slow (Fig. \ref{fig.saddlenode}) \cite{Tufcea2015}. This feature has been termed a "dynamical ghost" in recognition that a vestige of the steady state is present in the dynamics of the system near the bifurcation point \cite{Strogatz2014}.
   
To illustrate the effect on patterning time imposed by the regulatory architecture of the bistable switch, we compared it with a regulatory motif which lacks the cross repressive feedback between $A$ and $B$ and instead relies on strongly cooperative activation of $A$ by $M$. In this non-feedback case, there is only one steady state and the expression level of $A$ has a sigmoidal dependence on the level of morphogen, so $A$ appears absent for low morphogen values, while $A$ is expressed for high morphogen values. We defined this model to generate an outcome that resembles the bistable switch case (see \nameref{S1_Text}), although quantitative comparison with the feedback case is limited by the specific parameters chosen. Nevertheless, the results suggest that, in the absence of feedback, patterning time is in general faster than the bistable switch for all values of the signal. Importantly, since the stable state never loses stability there is no dramatic increase in patterning time close to the boundary $M_A$ (Fig. \ref{fig.patterningtime}).  

The comparison in dynamical response of the bistable switch to the non-feedback motif can be extended to changes in parameters $\alpha$ and $\delta$, the production and degradation rates respectively. Concretely, keeping the ratio $\alpha/\delta$ fixed, the response speed of gene $A$ over $B$ can be controlled. In both cases, reducing the time scale response of $B$ speeds up the whole transient and reduces the patterning time (Fig. \ref{fig.patterningtime}). In the case of the non-feedback model the reduction of patterning time is linear with respect to changes in $\alpha$ and $\delta$. By contrast, the patterning time of the bistable switch is much more resilient to changes in $\alpha$ and $\delta$ (Fig. \ref{fig.patterningtime}). This feature provides some robustness in patterning time to changes in parameters and means that the bistable switch is intrinsically slower than the non-feedback mechanism. 

\subsection*{Intrinsic fluctuations accelerate cell fate change in the monostable zone}

\begin{figure}[t]
\centering
\includegraphics[width=0.9\columnwidth]{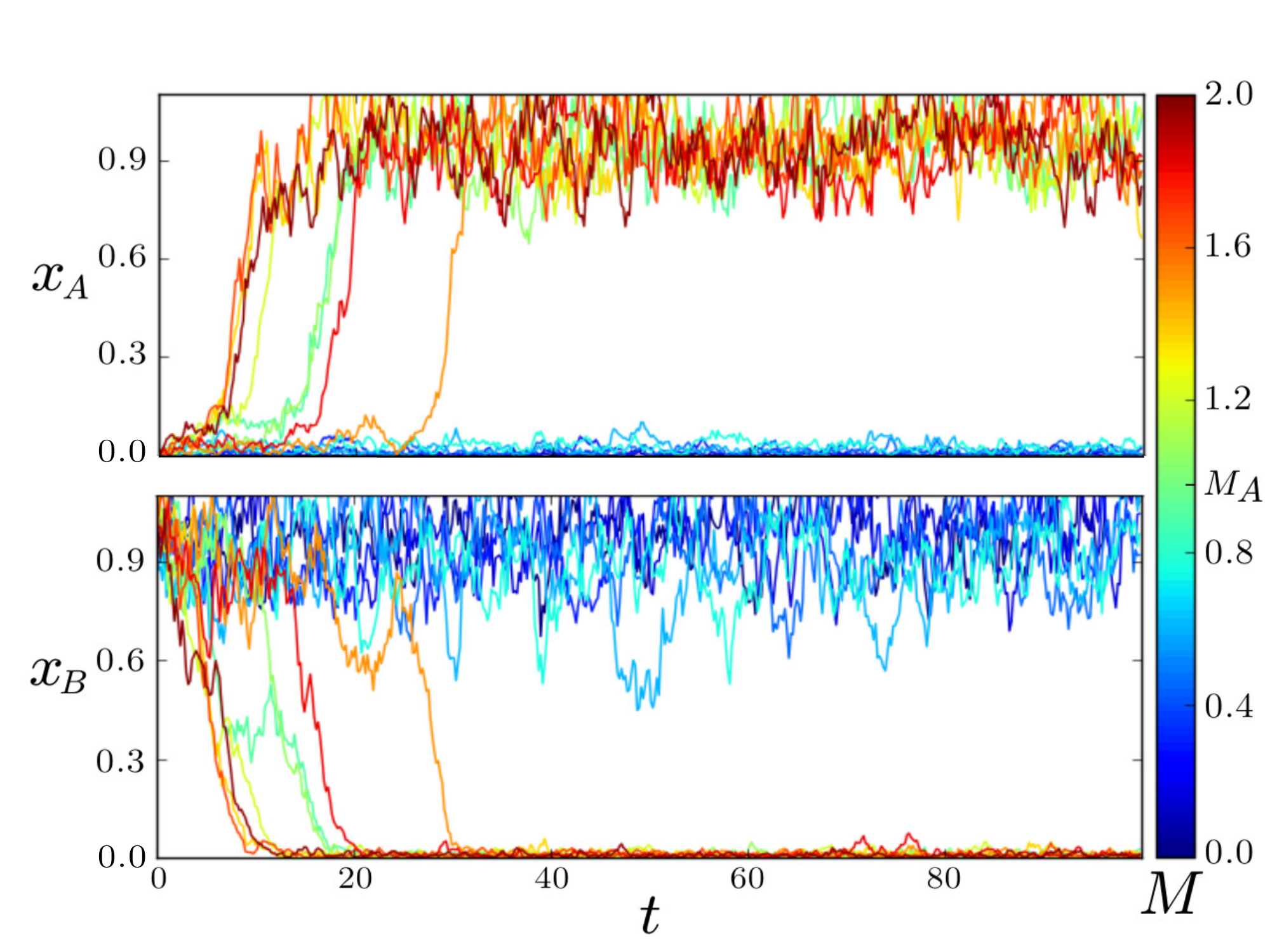}
\caption{ \label{fig.transientprofilenoise} \textbf{ Transient stochastic profiles to the steady state for different morphogen values.} Each line correspond to one CLE realisation with a different signal input. Parameters and ranges correspond with those of Fig. \ref{fig.transientprofile} with $\nu_A=\nu_B=1$ and $\Omega=100$.}
\end{figure}

The inherent fluctuations in the biochemical events that control gene expression, such as production and degradation events, introduce stochasticity into the expression dynamics (\emph{e.g.} \cite{Kepler2001,Swain2002}). This can be modelled mathematically using a variety of techniques. The stochastic evolution of gene expression can be obtained numerically by reproducing the kinetic reaction system in Eq.(\ref{eq.reaction}) using Gillespie simulations, or by integrating the approximate CLE (Eq. (\ref{eq.CLE})) (see \nameref{S1_Text}). The resulting gene expression trajectories introduce variability not only in the transient genetic expression, but also in the average patterning times (Fig. \ref{fig.transientprofilenoise}). 

\begin{figure}[ht!]
\centering
\includegraphics[width=0.9\columnwidth]{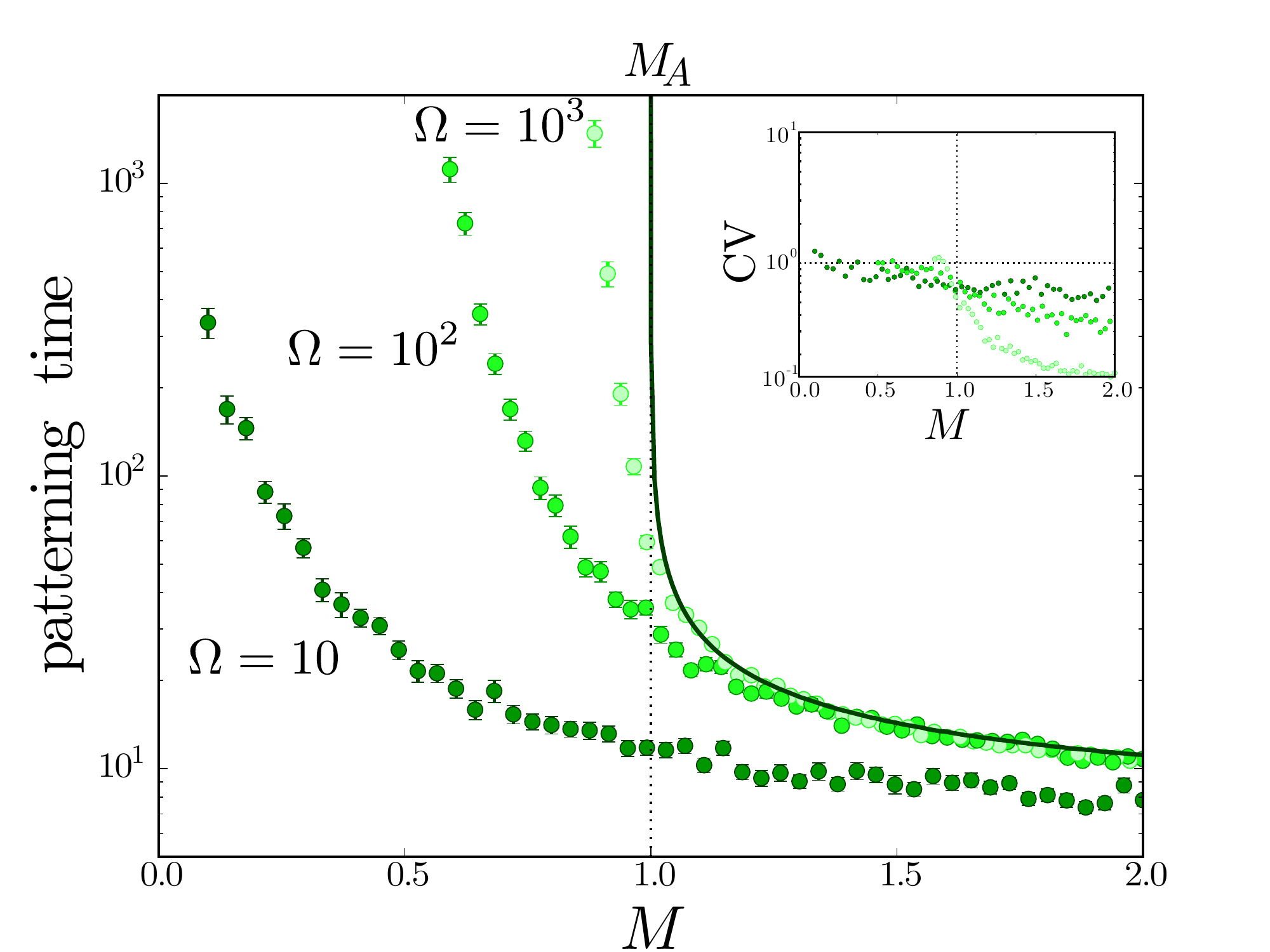}
\caption{\label{fig.signalboundaryOS} \textbf{Intrinsic noise changes average patterning time along the tissue}. Patterning times are compared for different values of $\Omega$ (circles) and with the deterministic behaviour (line). Each point corresponds to the average mean first passage time of 100 CLE realisations. Error bars correspond with standard error of the mean. Inset: Coefficient of variation of the patterning time. Parameters are the same as in Fig. \ref{fig.transientprofilenoise}.}
\end{figure}

We first focused on how patterning time is altered in the monostable zone where $M>M_A$. Comparison with the deterministic simulations indicate that the average patterning times for large values of the signal ($M\gg M_A$) are not affected by intrinsic fluctuations. By contrast, however, for values close to $M_A$, the patterning time is markedly lower in the stochastic simulations compared to the deterministic model (Fig. \ref{fig.signalboundaryOS}). Close to the threshold $M_A$, the slow dynamics introduced by the saddle-node bifurcation can not detain the expression dynamics because the intrinsic noise allows the system to explore the dynamical landscape and hence escape from the dynamical ghost. This accelerates the transient towards the final state while keeping transient expression levels similar to the deterministic ones.  Thus stochasticity in gene expression provides a mechanism that counterbalances the slow dynamics associated with the saddle-node bifurcation present in the deterministic system.

We next examined the effects of stochastic fluctuations in the bistable zone. Close to the bifurcation, the stable steady state B is marginally stable and intrinsic noise can result in sufficient repression of gene $B$ and activation of gene $A$ to switch to the stable steady state A. This stochastic switching occurs via a similar trajectory in gene expression space and on the same time scale as the patterning time in the monostable zone, resulting in an average patterning time that varies continuously between the monostable and bistable zones (Fig. \ref{fig.signalboundaryOS}). Consequently the resulting pattern boundary is no longer located at $M\simeq M_A$ but shifts inside the bistable zone. This behaviour is inherent to the bistable switch and is not found in the non-feedback circuit where noise cannot change the position of the pattern boundary, which is always located at $M\simeq M_A$ (\nameref{fig.signalboundarySTO}).  

\subsection*{Stochastic switching positions the pattern boundary inside the bistable zone}

Throughout the bistable region of the system, sufficiently large fluctuations will result in spontaneous switching. Such a transition is always possible, with a patterning time that increases super-exponentially as the signal level $M$ decreases (Fig. \ref{fig.signalboundaryOS}). Such a marked increase in patterning time can lead to biologically unfeasible switching times for a range of signal levels. In this range, the system will be resilient to intrinsic noise. By contrast, when the time scale of noise-driven switching is comparable to the patterning time of the tissue, the position and precision of the pattern boundary will be altered. Moreover, since the change in patterning time occurs continuously across the bistability threshold $M_A$, stochastic transitions will always be relevant during tissue patterning. 

For time scales larger than the degradation rates, the protein expression for each steady state will follow a certain probability distribution around the deterministic steady state. The larger the typical number of molecules defining each steady state ($N_i=x_i\Omega$), the smaller the effect of fluctuations, the narrower the deviations from the deterministic phenotype, and the longer the typical switching times (Fig. \ref{fig.signalboundaryOS} and \nameref{fig.signalboundaryOmegaOS}). The tails of this distribution will determine the stochastic switchings. In this context, large deviation theory predicts an exponential dependence of the average stochastic switching patterning time on $\Omega$ \cite{Freidlin1998,Touchette2009}

\begin{equation}
T_s=C\mathrm{e}^{\Omega \mathcal{S}}, \label{eq.MAPtime}
\end{equation}
where $C$ is a prefactor to the exponential behaviour and $\mathcal{S}$ is the action of the stochastic transition. Equation (\ref{eq.MAPtime}) can be related to the Arrhenius law where $\Omega^{-1}$ plays the role of temperature, controlling the fluctuations of the expression state, and $\mathcal{S}$ plays the role of the activation energy of the transition between different states. Consistent with this, $\mathcal{S}$ changes with the level of the morphogen signal and provides a correlate of the dynamical landscape.

In one-dimensional cases, such as the auto-activating bistable switch, the dependence on the signal of the stochastic switching time can be obtained directly from knowledge of the probability distribution in a closed integral form \cite{Frigola2012,Jaruszewicz2013,Weber2013}. In multidimensional cases, such as the one we are studying, there is more than one path across the dynamical landscape linking the steady states and finding the values of $\mathcal{S}$ requires us to consider the contribution of the different paths. Each path $\varphi_\tau$ of duration $\tau$ has a different probability that can be written as \cite{Freidlin1998,Touchette2009} 

\begin{equation}
P(\varphi_\tau)\sim \mathrm{e}^{-\Omega \mathcal S(\varphi_\tau)}, \label{eq.pathprob}
\end{equation}
where the exponential dependence on $\Omega$ predicts that for large enough numbers of proteins, the stochastic switching process will occur following the neighbourhood of the path $\varphi^*$ that minimises the action,
\begin{equation}
\mathcal S\equiv \mathcal S(\varphi^*)=\min_{\tau,\varphi_\tau} \mathcal S(\varphi_\tau). \label{eq.Smin}
\end{equation}

Applying the Eikonal exponential dependence of Eq. (\ref{eq.pathprob}) to the CLE describing the stochastic dynamics of the bistable switch (Eq. (\ref{eq.CLE})), a closed expression for the action can be obtained describing the stochastic switching process that for a general CLE of the form $\dot x=f(x)+g(x)\xi(t)$ gives \cite{Freidlin1998},
\begin{equation}
\mathcal{S}(\varphi_\tau)=\frac{1}{2}\int_{0}^{\tau} \left\Vert \dot\varphi_\tau(t)-f(\varphi_\tau(t)) \right\Vert^2_{g(\varphi_{\tau}(t))}\mathrm d t\label{eq.action}
\end{equation}
where $f(\varphi_\tau)$ is the deterministic field that describes the phenotypic landscape and from eq. (\ref{eq.CLE})  is,
\begin{equation}
\left(\begin{array}{c}f_A(x_A,x_B)\\f_B(x_A,x_B)\end{array}\right)=\left(\begin{array}{c} p_A(M,x_B)-x_A\\\alpha p_B(x_A)-\delta x_B  \end{array}\right)\label{eq.field}
\end{equation}
and the norm $\Vert\bullet\Vert_{g(\varphi_\tau)}^2$ in (\ref{eq.action}) corresponds with the inner product $\left\langle \bullet,\left(g(\varphi_\tau)g(\varphi_\tau)^\top)\right)^{-1}\bullet  \right\rangle$, where $g(\varphi_{\tau})g^{\top}(\varphi_{\tau})\equiv D$ is the diffusion tensor given by the noise intensity that for eq. (\ref{eq.CLE}) reads,
\begin{equation}
D(x_A,x_B)=\left(\begin{array}{cc} \nu_Ap_A(M,x_B)+x_A&0\\0&\nu_B\alpha p_B(x_A)+\delta x_B \end{array}\right).\label{eq.difftens}
\end{equation}

Thus, the value of the action can be obtained by the numerical minimisation Eq. (\ref{eq.Smin}) of the action functional Eq. (\ref{eq.action}) (see \nameref{S1_Text}). The result of this minimisation gives both the rate of stochastic switching for different values of the signal (\ref{eq.MAPtime}) and the minimum action path (MAP) describing the transient expression profiles of $A$ and $B$ during the switching. 

\begin{figure}[ht!]
\centering
\includegraphics[width=0.62 \columnwidth]{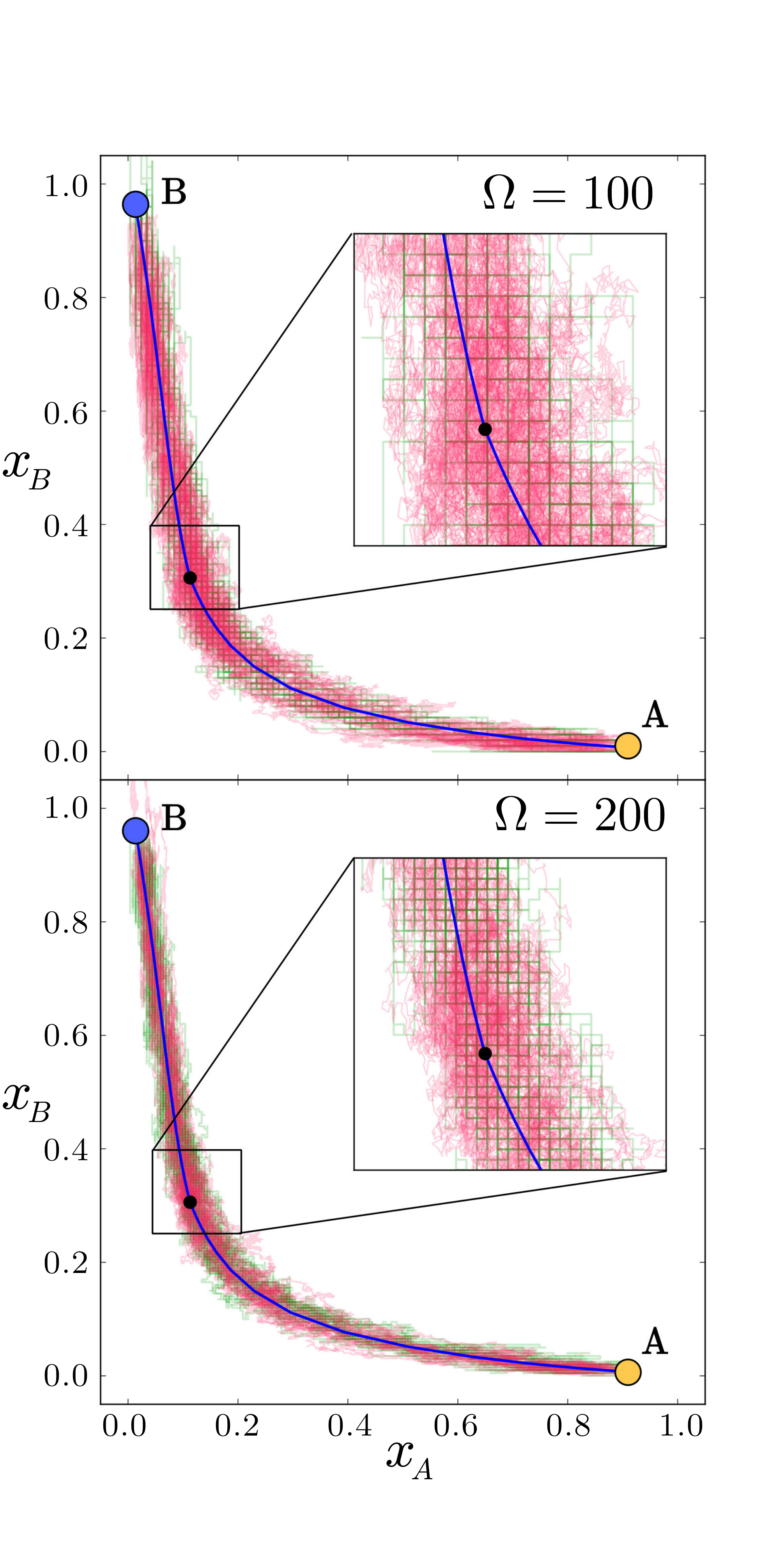}
\caption{ \label{fig.MAPnormal} \textbf{Stochastic switching trajectories concentrate along the MAP.} Stochastic switching trajectories are compared for CLE (red) and exact realisations of the kinetic reactions through the Gillespie algorithm (green) for two different values of $\Omega$ for the stochastic transition $A\rightarrow B$. The trajectories concentrate along the MAP (solid blue line), which passes through the saddle point (black circle). Parameters are those of Fig. \ref{fig.transientprofilenoise}. }
\end{figure}

To test the validity of the action minimisation, we compared the MAP with the stochastic switching trajectories resulting from simulations of the kinetic reaction scheme (\ref{eq.reaction}) and the CLE (\ref{eq.CLE}). This confirmed that stochastic trajectories concentrate around the MAP with increasing $\Omega$ (Fig. \ref{fig.MAPnormal}). This supports the validity of the MAP framework for estimating the trajectory of the transition and the computational efficiency, compared to stochastic simulations, makes it a useful complement to other techniques. Notably, the trajectory predicted by the MAP is distinct from the deterministic steepest-descent through the dynamical landscape given by the deterministic equations. The resulting path is shaped by the changes in intrinsic noise for different expression states through $g(\varphi)$. Thus the MAP provides the means to explore the consequences of stochastic mechanisms, such as expression bursts, that are unavailable in deterministic descriptions. 

\begin{figure}[h]
\centering
\includegraphics[width=0.9 \columnwidth]{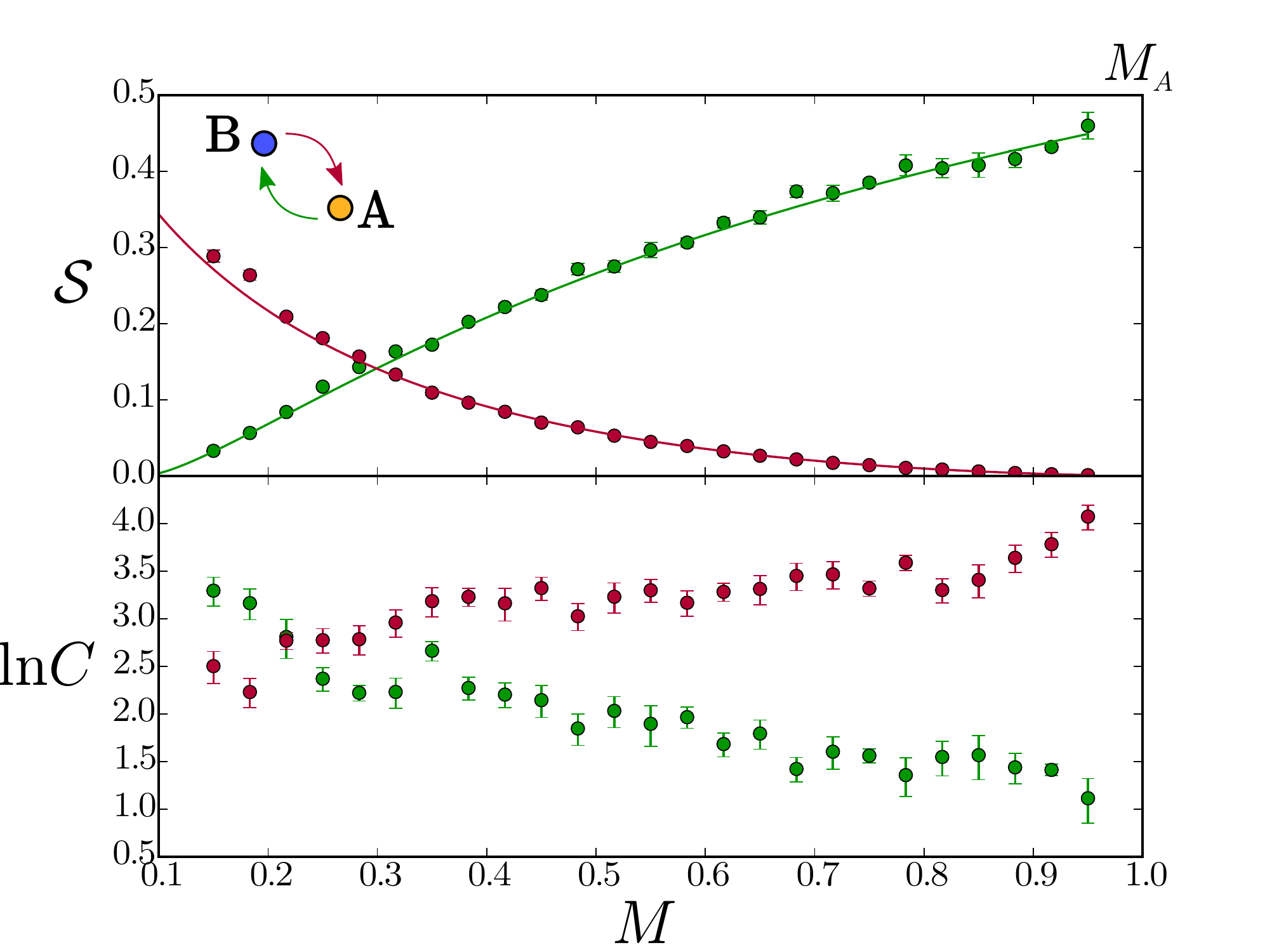}
\caption{\label{MFPTaction} \textbf{Stochastic switching rates change along the bistable zone.} Dependence of the action $\mathcal{S}$ and the prefactor $C$ on the signal, comparing the minimisation of the action functional (solid lines) with the resulting patterning times for 10000 averaged switching CLE trajectories for each signal value (circles). Error bars indicate standard error from the fitting (see \nameref{S1_Text}). Parameters are those of Fig. \ref{fig.transientprofilenoise}.}
\end{figure}

In addition to the MAP, the validity of the action can be tested by comparing the switching times obtained from CLE stochastic realisations with the exponential dependence of switching time obtained from the action (\ref{eq.MAPtime}) (Fig. \ref{MFPTaction}). This shows that the action allows patterning time to be determined with logarithmic precision for sufficiently large values of $\Omega$ \emph{i.e.} $\ln T_s=\ln C+\Omega \mathcal{S}\simeq\Omega \mathcal{S}$. Such values of $\Omega$ should involve patterning times greater than the patterning time in the monostable zone (Fig. \ref{MFPTaction}). This reduces the necessity to determine the prefactor $C$, which is not given by the action minimisation. In cases where $\Omega$ is not large enough, the prefactor has a relevant contribution and can be obtained with the help of the minimised action, from a reduced number of CLE simulations (see \nameref{S1_Text}). 

To characterise the effect of stochastic switching in the bistable zone, it is necessary to tally the change of fate $B\rightarrow A$ with its opposite $A\rightarrow B$ (Fig. \ref{MFPTaction}). The shape of the MAP and the values of the action for transitions in both directions ($\mathcal S_{AB}$ and $\mathcal S_{BA}$) change along the bistable zone (Figs. \ref{MFPTaction} and \nameref{fig.MAPsignal}): transitions from A to B become less probable, and B to A more probable as $M$ increases. This is translated as opposite trends of $\mathcal S_{BA}$ and $\mathcal S_{AB}$ with $M$. As a result, the residence time of the two states become equal ($\mathcal{S}_{BA}\simeq\mathcal{S}_{AB}$) at an intermediate value of the signal $M\simeq 0.3$. This predicts a new steady state position for the pattern boundary in the bistable zone. Away from this signal, the rate of one of the stochastic transitions becomes small compared with the other, resulting in one of the two states dominating. Strikingly, the location of the steady state boundary, whilst not dependent on $\Omega$, is different to that predicted by the deterministic system at $M=M_A=1.0$.

\subsection*{Expression bursts shape the stochastic switching process}

One contributor to the stochastic nature of gene expression is the inherent pulsatility of transcription/translation -- so-called 'bursty' expression -- which results in sporadic interspersed periods of expression and quiescence \cite{Blake2003,Raj2006,Pare2009,Thattai2001}. The framework we have developed allows us to explore the effect of the size of these bursts of expression, $\nu_A$ and $\nu_B$, on the behaviour of the system whilst keeping the deterministic dynamical landscape constant. The action is independent of $\Omega$ but is dependent on the burst sizes $\nu_A$ and $\nu_B$, which hence alter the MAP. The MAP approach is therefore suitable for capturing the differences in the effects of different noise sources. 

\begin{figure}[ht!]
\centering
\includegraphics[width=0.77 \columnwidth]{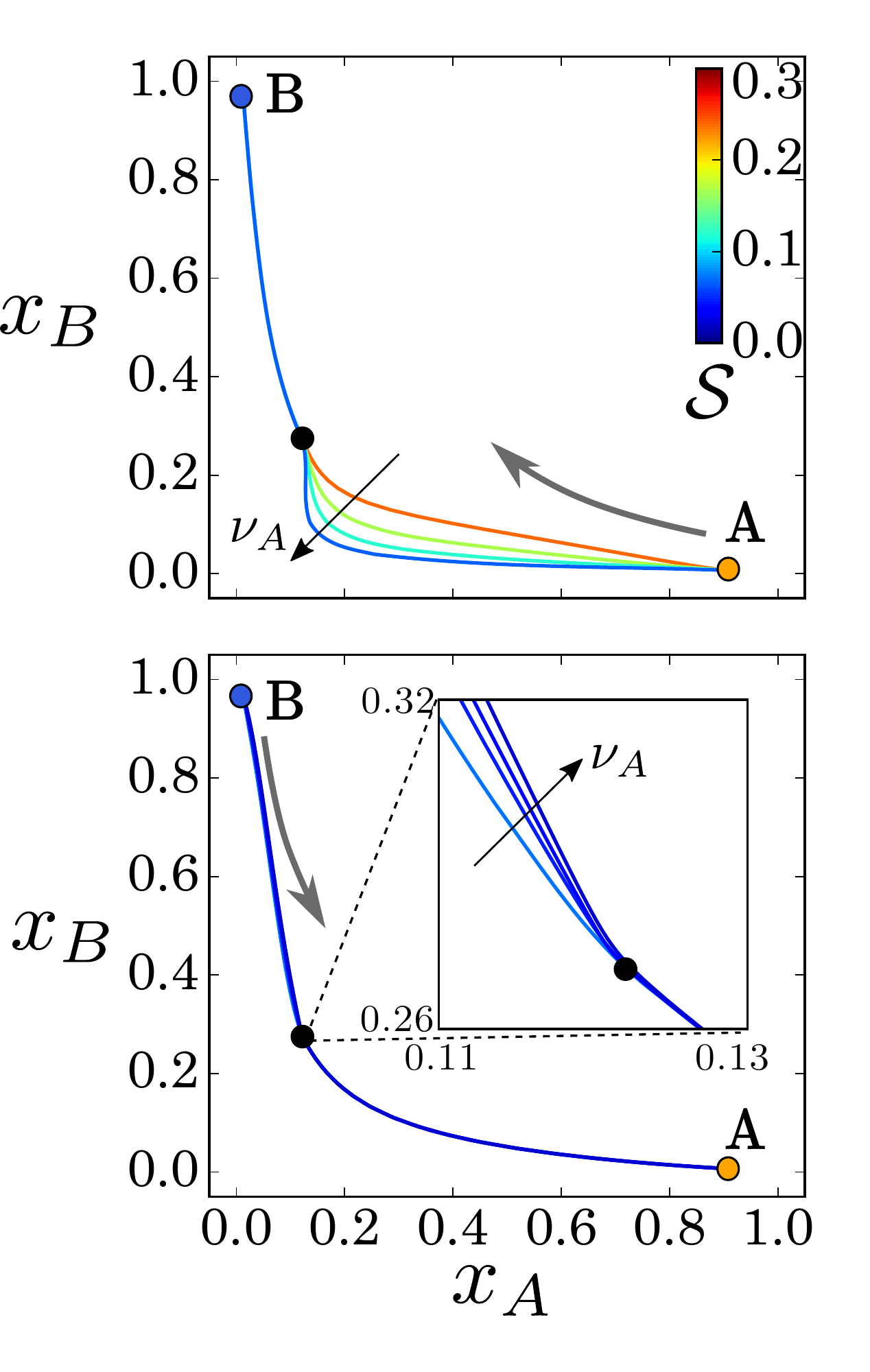}
\caption{\label{MAPnuA}\textbf{Stochastic switching trajectory and rate change with the burst size $\nu_A$.} Change of MAP for different values of the relative burst size $\nu_A=1,3,5,10$ for both switching processes $A\rightarrow B$ (top) and $B\rightarrow A$ (bottom). The value of the action (line color) changes with $\nu_A$ . The position of the different steady expression states are marked with color circles for state A (orange) and B (blue) as well as the saddle points (black). Morphogen signal is  $M$=0.45. Parameters are those of Fig. \ref{fig.transientprofilenoise}.}
\end{figure}

Intuitively, a larger burst size will introduce more noise in the expression of a protein and therefore facilitate a transition. This is confirmed quantitatively by comparing the actions of both switching processes (Figs. \ref{MAPnuA} and \nameref{MAPnuB}), which reveals a reduction in the actions as $\nu_A$ and $\nu_B$ increase. In the particular case of Fig. \ref{MAPnuA}, the reduction in action for the transition A to B is much greater than the reduction in action for the reverse transition. Whether changes in burst size can modify the directionality of transitions therefore warranted investigation.

Perturbations in the shape of the MAP are also evident (Fig. \ref{MAPnuA} and \nameref{MAPnuB}), suggesting that the greater the burst size of $A$ the less activation of $B$ is necessary for the repression of A. Equivalently, the greater the burst size of $B$, the less activation of A is necessary to repress B. This suggests that changes in $\nu_A$ and $\nu_B$ produce contrasting effects on the genetic profiles during switching. By contrast, a homogeneous increase in the noise, through a reduction in the typical number of molecules $\Omega$, does not result in changes in the switching path. 

\begin{figure}[ht!]
\centering
\includegraphics[width=0.83\columnwidth]{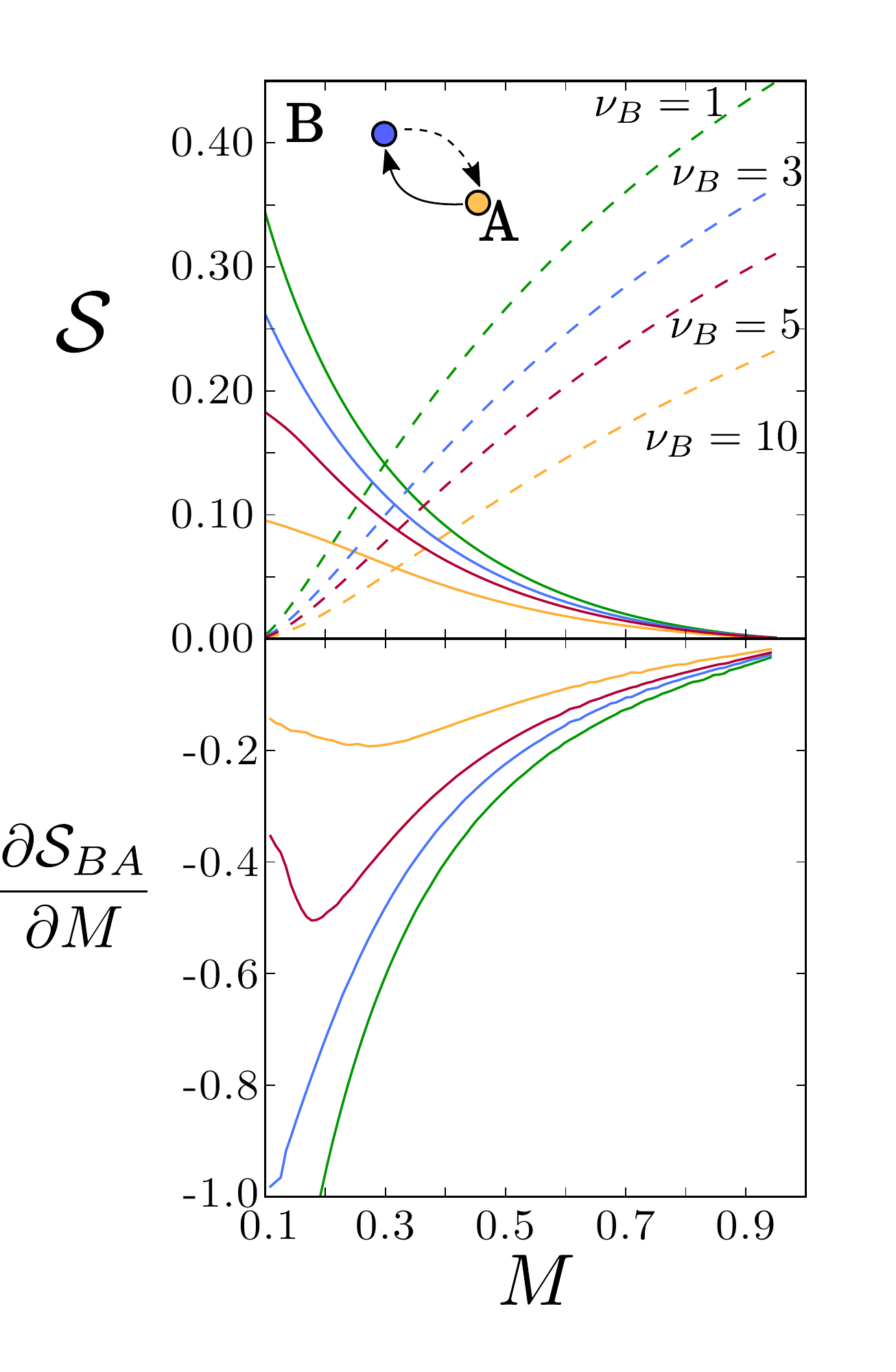}
\caption{\label{fig.actionsnuB} \textbf{Burst size modifies the change in action along the tissue.} The action along the tissue is evaluated for different values of $\nu_B$ for both switching transitions: $B\rightarrow A$ (solid line) and $A\rightarrow B$ (dashed lines), and its derivative over the morphogen concentration (bottom panel). Increasing the morphogen biases the cells towards state A. Parameters are those of Fig. \ref{fig.transientprofilenoise}.}
\end{figure}

A more detailed analysis of the effect of the burst sizes on the action profiles reveals that the approximate effect of an increase in burst size is to reduce both actions by a factor which is homogeneous across the tissue (Figs. \ref{fig.actionsnuB} and \nameref{fig.actionsnuA}). This has the effect that, where $\mathcal S_{BA}$ is larger than $\mathcal S_{AB}$, the reduction in $\mathcal \mathcal S_{BA}$ as burst size increases is greater. By contrast,  increases in burst size result in greater reduction in $\mathcal S_{AB}$  where $\mathcal S_{AB}$ is larger. Therefore, curiously, the burst parameters do not affect the directionality of the transitions, since they reduce the actions by the same amount at the steady state boundary position. At lower values of the morphogen signal, state B is still favoured, as at higher values is state A. Thus different burst sizes produce different action profiles along the tissue  (Fig. \ref{fig.actionsnuB}) and these will be translated into different transients and precisions of the boundary, but the steady state position of the boundary remains the same.

\subsection*{The gene expression boundary propagates through the tissue at a velocity determined by stochastic effects}

The action $\mathcal S_{BA}$ grows superlinearly as the signal decreases (Fig. \ref{MFPTaction}), and the stochastic switching time grows exponentially with $\Omega$. This results in patterning times (residence time at $B$ denoted by $T_B$) that grow super-exponentially as the signal decreases. A consequence of this is that patterning time varies dramatically, differing by orders of magnitude, along the bistable zone. This can result in switching times much larger than those relevant to biological processes, in which case the steady state would never be reached. Such a big difference allows the separation of the bistable zone into two regions at any time $t$ during the transient: an area where the stochastic switching time towards state A is much smaller than the current time, $T_B\ll t$, (this region expresses predominantly $A$); and an area where the switching time towards state A is much larger than the current time, $T_B\gg t$, (this region expresses predominantly $B$). The very large variation of $T_B$ along the gradient $M$ allows this separation, since the area of the tissue where $t\simeq T_B$ is small. At different times the boundary between these two behaviours will be located at different spatial positions. Specifically, the boundary will be located at positions corresponding with values of the signal where switching time coincides with the current time $t=T_B$, \emph{i.e.}  a value of the signal $M_s$ where $\mathcal S_{BA}=\frac{1}{\Omega}\ln (t/C)$. Thus, a bistable switch will produce a pattern boundary that advances away from the morphogen source with a velocity

\begin{eqnarray}
v(t)&=&\frac{\mathrm d M_s}{\mathrm d t}=\left(\left.\frac{\mathrm d T_B}{\mathrm d M}\right|_{M_s}\right)^{-1}=\nonumber\\
&=& \frac{1}{t} \left(\left. \frac{\partial \ln C}{\partial M}\right|_{M_s}+\Omega \left.\frac{\partial \mathcal S_{BA}} {\partial M}\right|_{M_s}  \right)^{-1}\simeq \frac{1}{\Omega t} \left(\left.\frac{\partial \mathcal S_{BA}} {\partial M}\right|_{M_s}\right)^{-1},   \label{eq.velocity}
\end{eqnarray} 
where, as described above, for large enough number of proteins (large $\Omega$), the result is independent of the prefactor $C$. The boundary velocity (\ref{eq.velocity}) can therefore be determined by the value of $\Omega$ and by the dependence of the action on the signal, which is readily computed numerically (see Fig. \ref{fig.actionsnuB}). The greater the typical number of proteins, the slower the advance will be and in the deterministic limit ($\Omega\rightarrow\infty$), the boundary velocity vanishes. 

\begin{figure}[h]
\centering
\includegraphics[width=0.9\columnwidth]{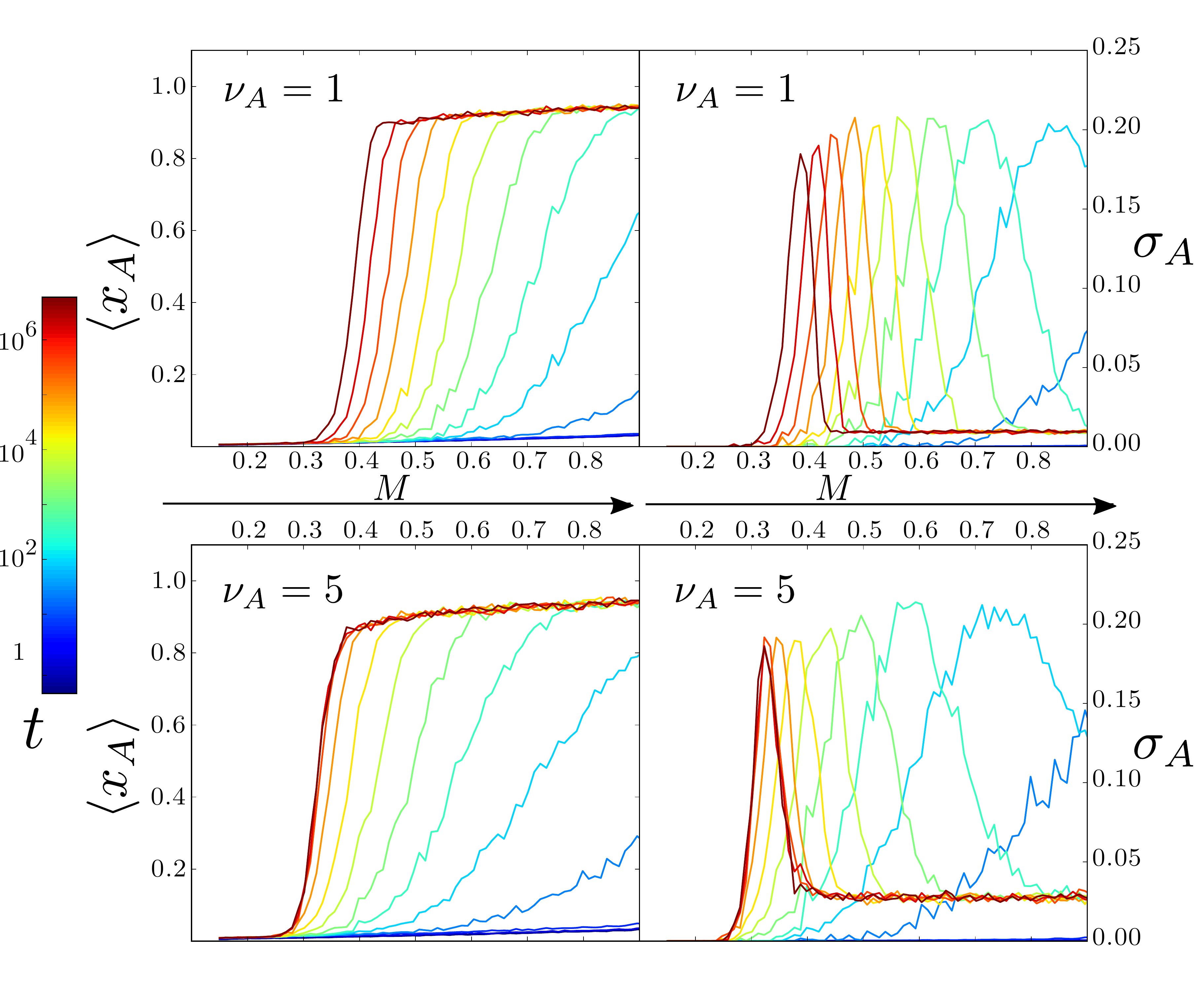}
\caption{\label{fig.patternshift} \textbf{Switching events result in a travelling boundary depending on stochastic parameters.} Mean and standard deviation in expression of the morphogen activated gene $A$ along the tissue at different time points for different burst sizes. Results correspond to averaging of 500 trajectories with $\Omega=120$; the rest of the parameters are the same as in Fig. \ref{fig.transientprofilenoise}.}
\end{figure}

\begin{figure}[h]
\centering
\includegraphics[width=1.0\columnwidth]{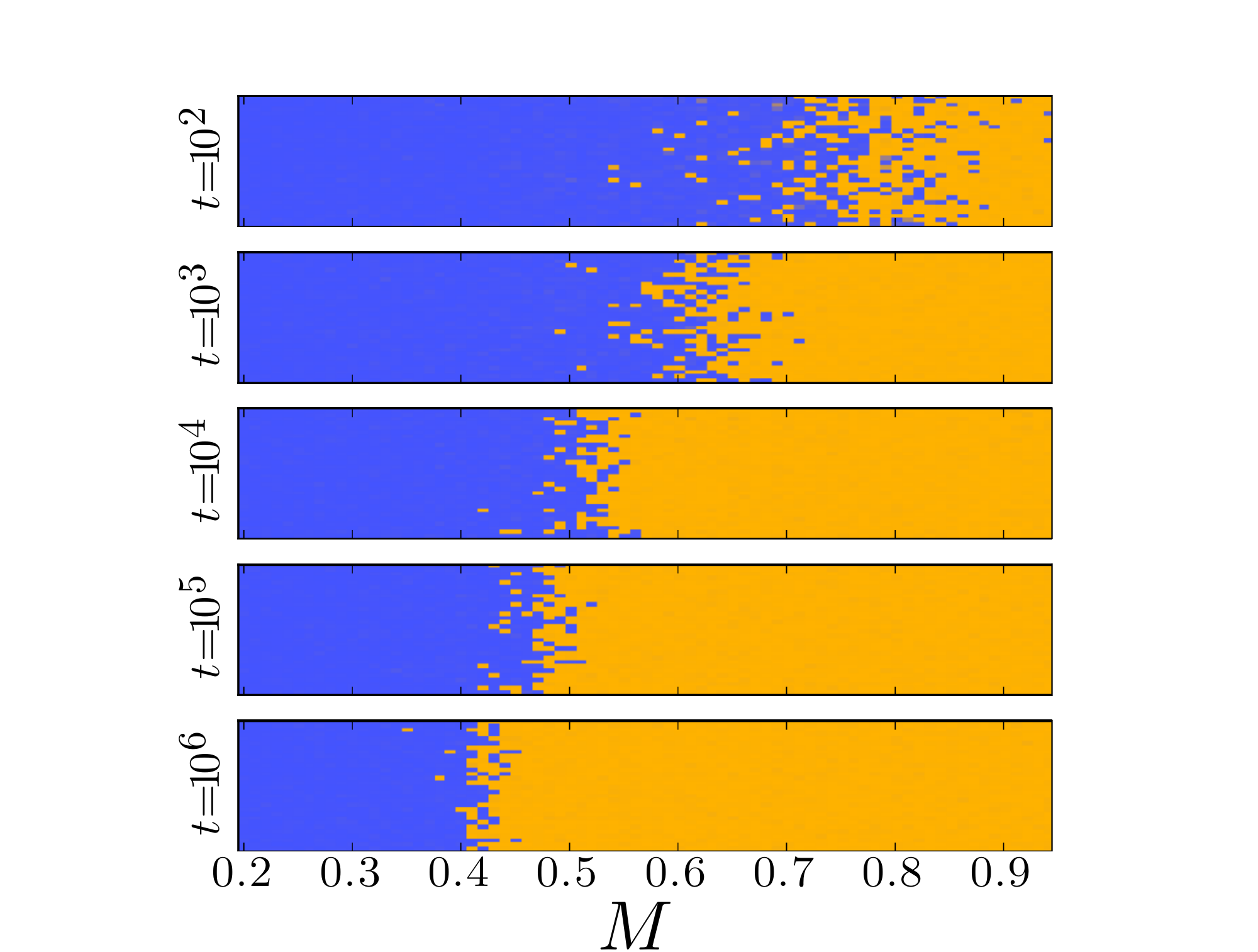}
\caption{\label{fig.patterntime} \textbf{Stochastic reproduction of the patterning process.} Patterning tissue reproduced as CLE trajectories at different time points. Each tissue is composed of $20\times 70$ cells for the same parameters used in Fig. \ref{fig.patternshift} top row ($\nu_A=1$). The colour of each cell is the RGB linear combination of blue and orange weighted with the expression states $x_A$ and $x_B$.}
\end{figure}

The velocity of the advancing boundary (\ref{eq.velocity}) decreases in time and the rate of this deceleration is given by the terms $1/t$ and $\frac{\partial \mathcal S_{BA}} {\partial M}$, where the absolute value of this second term increases as the boundary propagates. As a result, the boundary can appear static for short periods of time, only revealing its movement when tracked for several orders of magnitude in time (Figs. \ref{fig.patternshift} and \ref{fig.patterntime}). This is relevant for biological time scales, where a slowly travelling boundary may appear fixed. The position of the boundary at any time will be determined by $\Omega$ and the change of the action with the signal. 

The travelling boundary induced by a graded morphogen signal can also be observed in stochastic simulations implemented on an array of cells (Fig. \ref{fig.patternshift}). Moreover, for each value of signal at each time point, the standard deviation in $x_A$ between cells for each value of $M$ can be measured, this gives an estimate of boundary precision. As expected the maximum variation is at the boundary $M_s$ (Fig. \ref{fig.patternshift}). Even at this position, however, the majority of individual cells are 'decisive', residing in either an A or B state and relatively few cells are in undefined/transition states where $x_A\simeq x_B$ (Fig. \ref{fig.patterntime}). The decisiveness of individual cells is a consequence of the relatively rapid transition between states once initiated, compared with the stochastic switching time scale. Thus stochasticity results in a salt and pepper distribution of cell identities close to the boundary. The sharpness of the boundary during the transient at a time $t$ is determined by the size of the zone where $t\simeq T_B$, and will therefore increase its precision with $\Omega$ (\ref{eq.MAPtime}). This indicates a trade-off between the sharpness of the boundary and the velocity towards the steady state (\nameref{fig.velOmega}). The larger $\Omega$, the more precise the boundary, but the slower the velocity of the boundary along the patterning axis (\nameref{fig.velOmegaprofile}). 

Finally, we asked how the burst size $\nu_A$ affects the dynamics of this travelling wave. Since alterations in $\nu_A$ produce different transition profiles $\mathcal S_{BA}(M)$  (\nameref{fig.actionsnuA}), this will affect the velocity of transitions. Specifically, an increase in burst size is predicted to increase the boundary velocity (Fig.\ref{fig.patternshift}). As a result, even though different values of $\nu_A$ have the same steady state boundary $M\simeq 0.3$, they will have boundaries that travel along the patterning axis at different velocities. Strikingly, however, it appears  (Fig. \ref{fig.patternshift}), that the patterning precision, measured as the width of the region with elevated standard deviation in $x_A$ between cells, is unaffected by the burst size, when comparing the patterns at a given time. This contrasts with the effect of $\Omega$, where a smaller $\Omega$ always increases the width of the boundary at any time point in the transient (\nameref{fig.velOmegaprofile}). 

\section*{Discussion}

Bistable toggle switches are a regulatory mechanism employed in many biological systems (\emph{e.g.} \cite{Enver2009,Zhou2011,Shu2013,Manu2009,Balaskas2012}). In developing tissues, these switches are encoded by cross-repressing transcription factors and are often coupled with spatially graded extracellular morphogen inputs. This provides a mechanism to convert the spatially continuous information provided by the gradient into discrete, localized stripes of gene expression, with the morphogen input controlling the position of the boundary. To explore the effect of gene expression noise on the dynamics and output of this mechanism we used a suite of mathematical techniques, including numerical simulations of the kinetic reactions, Chemical Langevin Equations, and Minimum Action Path theory. This combination of approaches provided consistent and complementary results. In particular, the use of Minimum Action Path theory allowed essential properties of the stochastic dynamical system to be accurately estimated without the necessity for computationally expensive stochastic simulations. The results revealed that cell fate determination occurs at different times along the tissue resulting in a wave of patterning that propagated away from the morphogen source. Intrinsic noise determined both the velocity at which the gene expression boundary moved and its final position in the tissue. We discuss the implications of these results for developmental pattern formation. 

In order to understand the role of intrinsic noise, we first studied the behaviour of the deterministic system where, for a homogeneous tissue expressing $B$, a change in cellular fate towards state A requires a signal exceeding the threshold $M_A$. State A is not reached immediately but undergoes a transient with a speed dependent on the signal level $M$. A signature of the bistable switch is that, at the threshold $M_A$, the system undergoes a saddle-node bifurcation, implying that at values of $M$ just above $M_A$, the transient in the deterministic system is very slow \cite{Strogatz2014}. However, our analysis indicates that the slow dynamics of the deterministic bistable toggle switch are essentially eliminated by the introduction of stochasticity into gene expression. In this situation, the fluctuations in gene expression allow a system close to $M=M_A$ to escape spontaneously from this region of dynamical space, speeding up the cell identity determination in the monostable zone ($M>M_A$). 

Introducing intrinsic fluctuations in gene expression also result in spontaneous switching events between states $B$ to $A$ within the region of bistability  $(M_B<M<M_A)$. The rate of these transitions depends on the noise intensity, and can vary by orders of magnitude along the patterning axis.  Comparing the effects of stochasticity on the behaviour of a bistable switch in the mono- and bistable zones suggests a trade off between distinct demands on the system. An increase in noise (smaller $\Omega$) allows the system to reach its steady state more swiftly in the monostable zone ($M>M_A$). The consequence of this, however, is that the increased noise increases the rate of spontaneous switching in the bistable region and therefore decreases the precision of the gene expression boundary. The importance of speed versus precision will be dictated by the particular biological requirements of the developing tissue. It is possible that compromises can be found that balance these demands. In this context it is noteworthy that increases in burst sizes $\nu$ are predicted to accelerate pattern formation without the same deterioration in precision that results from decreasing the system size ($\Omega$), at least at similar time points. Thus enhancing the bursting behaviour of gene expression might be one way to increase the speed of patterning without losing precision. Alternatively, additional mechanisms might be employed to bypass any trade off. In situations in which a fast, precise pattern is required, differential cell adhesion, or other intercellular communication strategies, could be used to correct patterning mistakes \cite{Zhang2012,Sokolowski2012,Lander2013}. Alternatively a morphogen signal that varies in time could be exploited to avoid the dynamical ghost, increasing the speed at which the steady state is approached. For example, a signal that reaches its peak amplitude rapidly after ligand stimulation and then progressively decreases might effectively bypass or minimise the time the system spends in a dynamical ghost. In this respect, it is notable that two well studied morphogens, Tgf$\beta$ and Shh, have both been reported to display the type of adapting signaling dynamics that could be well suited to this task. \cite{Sorre2014,Balaskas2012}.

The rates of stochastic switching between the two cellular states change along the gradient with opposite trends. As a result, the steady state boundary for the stochastic system is located inside the bistable zone at the point where the rates of switching between the states balance. Strikingly, this position is distinct from the one defined by the deterministic system and does not change with increasing values of $\Omega$. The position is also robust to expression burst size, and is therefore determined by the parameters describing the deterministic switch. The duration of the transient towards this steady state is defined by the switching times and these vary super-exponentially along the tissue. This results in a traveling front of switching in which the pattern boundary moves away from the morphogen source. The velocity of the advancing boundary decreases dramatically as the boundary propagates. This slowing of the movement means that the gene expression pattern in a tissue might never reach steady state within a biologically realistic time scale. In this scenario the tissue patterning is always effectively pre-steady state, with a boundary moving so slowly it appears still, but in a different position from the steady state. The speed of boundary movement is affected by stochastic parameters such as the typical number of proteins $\Omega$ or expression bursts size $\nu$ that increase the velocity of boundary propagation even though the nominal steady state position remains constant. Taken together therefore, the intrinsic fluctuations in gene expression qualitatively alter the timing and positioning of the boundary, identifying a prominent role for noisy gene expression in pattern formation.

To explore the properties of the stochastic switching, the minimisation of the action proved an efficient strategy. It provides a much more computationally efficient means to gather information about the stochastic properties of the system than conventional CLE or Gillespie integrations (see \nameref{S1_Text}). In the current study, the action was computed for the CLE (\ref{eq.CLE}). Action minimisation approaches can be extended to more complex systems such as the full kinetic reaction scheme, or include, for example, the dynamics of mRNA production and decay and discrete promoter states \cite{Newby2011,Lv2012,Lv2014}. In cases in which the number of proteins is very small and the Eikonal assumption (\ref{eq.MAPtime}) fails, further expansion in terms of powers of $\Omega$ can be used and large deviation principles still hold \cite{Roma2005}. The principal caveat of action minimisation strategies is the inability to find the prefactor $C$ \cite{Lv2014}. Nevertheless, even without the prefactor, the action allows time scales to be calculated with logarithmic precision and in many cases this is likely to be sufficient to identify the main patterning properties of biological tissues. For example we show that the dependence of the action on the level of morphogen allows the estimation of the velocity at which the pattern boundary moves.

The MAP approach also offers quantitative insight into the transient expression profiles of $A$ and $B$ during the stochastic switching between cell identities. The trajectory predicted by the MAP is distinct from that predicted by the steepest-descent of the deterministic landscape and is shaped by the burst size parameters (although not the system size parameter $\Omega$). Stochastic simulations were consistent with the MAP, indicating that gene expression during individual cell switching can be predicted even when high levels of stochasticity are introduced. In this view, the dynamical landscape imposed by the regulatory interactions ensures that as cells transit between states they pass close to the saddle point in gene expression space. In addition, the transient path is similar in the mono- and bistable region of the system. This suggests that, in developing tissues, changes in cell identity should be characterized by defined trajectories in the levels of expression of the key genes. Thus even in the presence of gene expression noise, tightly constrained paths between the cell states would be representative of underlying kinetic mechanisms driving the cell state transition. This is consistent with ideas of developmental canalisation \cite{Waddington1942,Huang2010} and leads to experimentally testable predictions that could be assessed using single cell resolution imaging of reporters for the two genes comprising the bistable switch. Observing coordinated changes in gene expression that matched MAP predictions would support the validity of the MAP approach and allow key parameters of the biological system to be estimated. Alternatively, even in the absence of live imaging data, it is possible that static data from batches of cells, using techniques such as single cell transcriptomics, might provide experimental tests of the approach. Experimental corroboration of gene expression dynamics characteristic of those predicted by the MAP would provide insight into the mechanisms controlling cell decisions and greatly strengthen the evidence underpinning the use of dynamical systems theory to model these developmental events.

A practical feature of MAP theory is that it can be extended to include more complex networks comprising several genes per cellular state. This would offer insight not only into simple binary decisions, such as that described in this study, but also provide a way to study transitions in more complex and realistic models of cell development \cite{Wells2015,Li2013}. Thus MAP theory has the potential to provide a powerful framework to explore and understand the role of noise and dynamics in cell state transitions during normal development and also in other situations such as artificial directed reprogramming experiments \cite{Zhou2011,Wang2014,Li2013}. More generally, our results emphasise that stochastic fluctuations in gene expression can influence the dynamics and outcome of gene regulatory networks and highlight the importance of developing the mathematical tools to explore these aspects of developmental patterning. 

\section*{Acknowledgments}
We would like to thank A. Kicheva and A. Sagner for many helpful comments on the manuscript.

\nolinenumbers

%
%
%
%


\clearpage
\section*{Supporting Information}

\subsection*{S1 Text}
\label{S1_Text}
\section*{Patterning without positive feedback loop $M\rightarrow A$}

In order to get a step-like sigmoidal response to the signal for a monostable system, it is necessary to use high cooperativity for the morphogen activation. In the context of the promoter state using the thermodynamical description this can be introduced as $h$ independent binding sites for the activating signal on the promoter each one facilitating the binding of the RNAp when bound by the signal \cite{Bintu2005},

\begin{equation}
p_A^*(M)=\left(1\!+\!\rho_A\left(1+\frac{x_B}{K_B}\right)^2\!\!\left(\frac{1+M/K_M}{1+fM/K_M}\right)^h\right)^{-1}\!\!\!\!\!\!.
\end{equation}

\begin{figure}
\centering
\includegraphics[width=0.8\columnwidth]{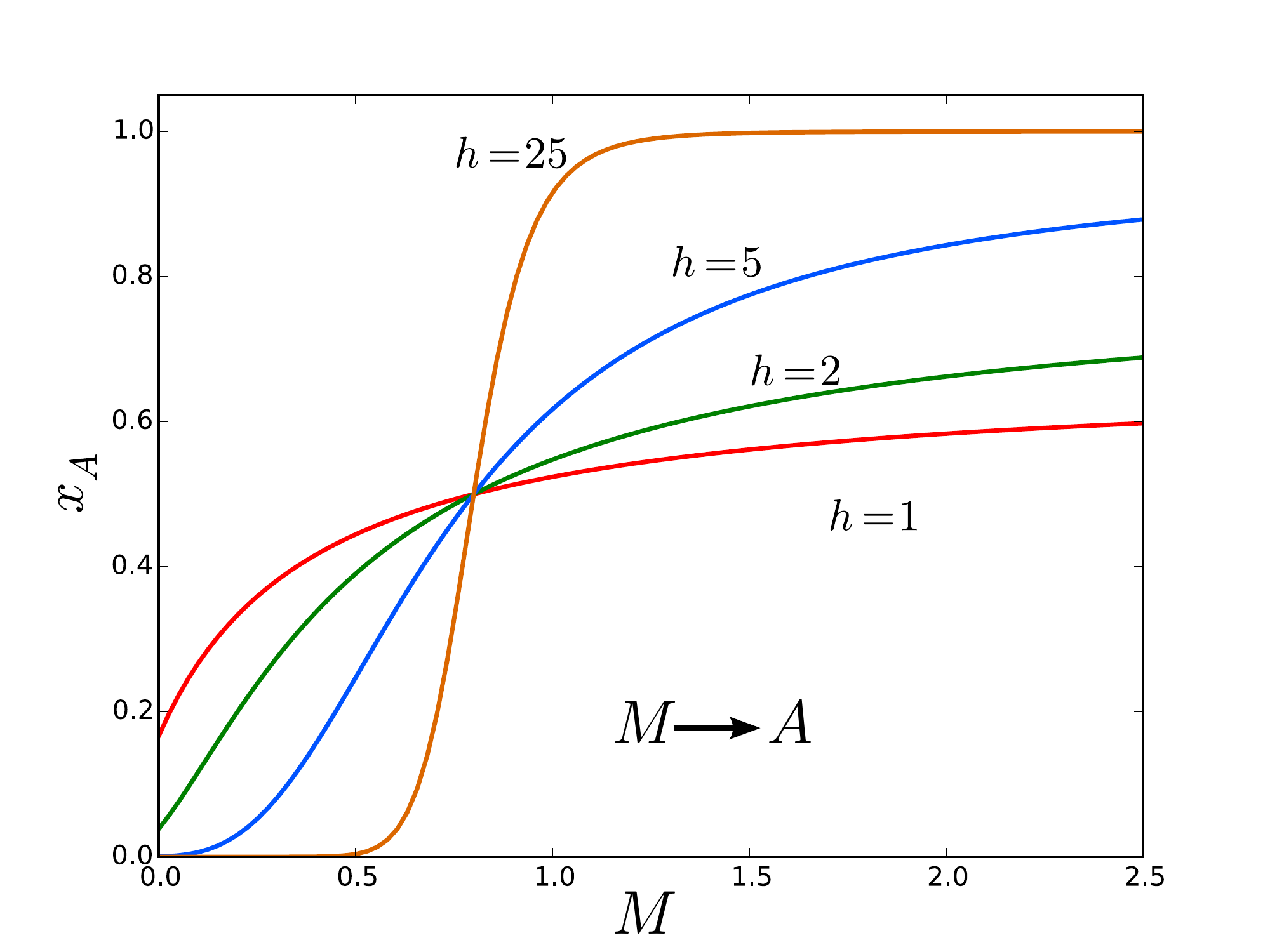}
\caption{\small \label{fig.mono}  Stationary expression of A without the positive feedback loop for different values of the cooperativity exponent $h$.}
\end{figure}

Even though the feedback loop is removed, a certain constant amount of $x_B$ is considered to represent the boundary position at the same signal value. Alternatively this is equivalent to redefining the value of $\rho_A$. For the comparison in the text, the values used are $h=25$ and $x_B=3 K_B=0.12$ which give a sharp boundary that reaches its maximum expression at $M\simeq 1$ (Fig. \ref{fig.mono}). The rest of the parameters are the same as the complete bistable switch to facilitate their comparison (see Fig. \ref{fig.landscape}).

The case without feedback cannot be directly compared with the bistable switch without a proper parameter exploration of both models. Nevertheless it is informative to determine how patterning time depends on the morphogen close to the boundary point $M=1.0$. For the case without positive feedback the expression dynamics are determined by the linear ODE,
\begin{equation}
\dot n_A=\alpha_A p_A^* - \delta_A n_A
\end{equation}
This predicts an exponential decay that for the initial condition $n_A=0$ is,
\begin{equation}
n_A(M,t)=\frac{\alpha_A p^*_A}{\delta_A}\left(1-\mathrm{e}^{-\delta_A t}\right)
\end{equation}

\section*{Numeric minimization of the Action}

The minimisation of the action (\ref{eq.Smin}) was performed by discretising the continuous path $\varphi_\tau$ in N straight segments between positions $\lbrace\phi_0,\phi_1,\phi_2,...\phi_N\rbrace$ of duration $\Delta t_i$ each. With this approximation, the action (\ref{eq.action}) can be written as,
\begin{equation}
\mathcal{S}(\varphi_\tau)=\frac{1}{2}\sum_{k=1}^{N}\sum_{i=\lbrace A,B \rbrace} \frac{\left(\dot\varphi^i_k-f_k^i \right)^2}{{D_k^i}}\Delta t_k, \label{eq.actiondiscrete}
\end{equation}  
where the index $i$ runs over the two genes and $k$ over each linear segment of the path. $D^i_k$ and $f^i _k$ are the $i$th component of the diagonal diffusion tensor (\ref{eq.difftens}) and the deterministic field (\ref{eq.field}) for the segment $k$ evaluated at its centre $x_k=(\phi_{k-1}+\phi_{k})/2$. Additionally, the velocity along a segment is considered constant $\dot \varphi_k=(\phi_{k}-\phi_{k-1})/\Delta t_k$. For a certain fixed total time $\tau$, the discretised action (\ref{eq.actiondiscrete}) can be minimised using a quasi-newtonian method with the help of the analytical expression for the derivative of the action on the direction of the $l$-th genetic component of segment point $s$,\begin{eqnarray}
\quad\qquad\frac{\partial \mathcal S}{\partial \phi_s^l}&=&\frac{f^l_{s}-\dot\varphi^l_{s}}{D^l_s}-\frac{f^l_{s-1}-\dot\varphi^l_{s-1}}{D^l_{s-1}}-\\
&-&\frac{1}{2}\sum_{j=\lbrace A,B\rbrace}\left[\frac{\dot\varphi^j_{s}-f^j_{s}}{D^j_s}f'^j_{l,{s}}\Delta t_{s}-\right.\nonumber\\
&-&\left. \frac{\dot\varphi^j_{s-1}-f^j_{s-1}}{D^j_{s-1}}f'^j_{l,{s-1}}\Delta t_{s-1} \right.+\nonumber\\
&+&\frac{(D')^j_{ls}}{2 (D^2)^j_{ls}}\left(\dot \varphi^j_s-f^j_s\right)^2\Delta t_s+\nonumber\\
&+&\left.\frac{(D')^j_{l,{s-1}}}{2 (D^2)^j_{l,{s-1}}}\left(\dot \varphi^j_{s-1}-f^j_{s-1}\right)^2\Delta t_{s-1}\right]\nonumber
\end{eqnarray}
where ${D'}^j_{l,s}$ and $f'^j_{l,s}$ are the derivative in the $l$ direction of the $j$th component of the diffusion tensor (\ref{eq.difftens}) and the  deterministic field  (\ref{eq.field}) of segment $k$. The minimisation of the action with $2\times N$ degrees of freedom was performed using the LFBGS minimisation strategy. The resulting value of the action $S_\tau$ is the minimum action for a certain path time $\tau$. The optimal action is found by minimising the resulting $S_\tau$ with $\tau$ by means of a simple  custom made line search method. Since the action here studied is a convex function of the possible paths, the minimisation process does not require an exploration of possible local minima. To speed up the convergence to the solution the starting path consists on two straight lines joining the initial and final phenotype with the saddle point. 

The actions for both switching directions $\mathcal S_{AB}$ and $\mathcal S_{BA}$ is determined by the boundary conditions of the integration fixing $\phi_0$ and $\phi_N$ at the initial and final expression amounts of the corresponding bistable states $A$ and $B$. In order to obtain the expression value of the bistable states for different morphogen signals, the bistable switch was reconstructed using continuation strategies through the \emph{PyDSTool} environment \cite{Clewley2012}.

\section*{Simulation of stochastic trajectories}

Simulation of the stochastic trajectories were performed using custom made code. Realisations of the Chemical Langevin Equation have been integrated using the Euler-Maruyama algorithm that reproduces the discrete Wiener process \cite{Maruyama1955,Ojalvo2012}, while the exact realisations of the kinetic reaction equations were performed using the Gillespie algorithm \cite{Gillespie1977}. Both methods require the reproduction of the whole trajectory and the computing time will be proportional to the trajectory time making these calculations very slow when computing stochastic switching events with increasing number of proteins $\Omega$. By contrast, the computation of the action is immediate, independent of the actual transition time between states, making it a useful tool for studying spontaneous cell fate change. 

The main drawback in computing the action is the unkown prefactor $C$ (\ref{eq.MAPtime}). This means that the transition time can be computed only to logarithmic precision \cite{Touchette2009,Lv2014}. In cases in which a greater precision is required, the prefactor $C$ can be obtained by fitting an exponential to average transition times obtained with the CLE or Gillespie at adequate values of $\Omega$, such that $\Omega$ is large enough that the behaviour can be considered exponential, and low enough to obtain reasonable computation times. This range can be guessed from the values of the action $\mathcal S$ while checking the goodness of the exponential fit. For the case of Fig. \ref{MFPTaction} the prefactor $C$ was computed by choosing a range where the mean first passage time depended exponentially on $\Omega$, and performing a least squares fitting to an exponential function $T=C\mathrm{e}^{\Omega \mathcal S}$. The least squares fitting was not linearized to give more weight to points of large $\Omega$, where the exponential behaviour is assured.

\clearpage
\subsection*{S1 Fig}
\label{fig.signalboundarySTO}
\begin{figure*}[ht!]
\centering
\includegraphics[width=0.5\textwidth]{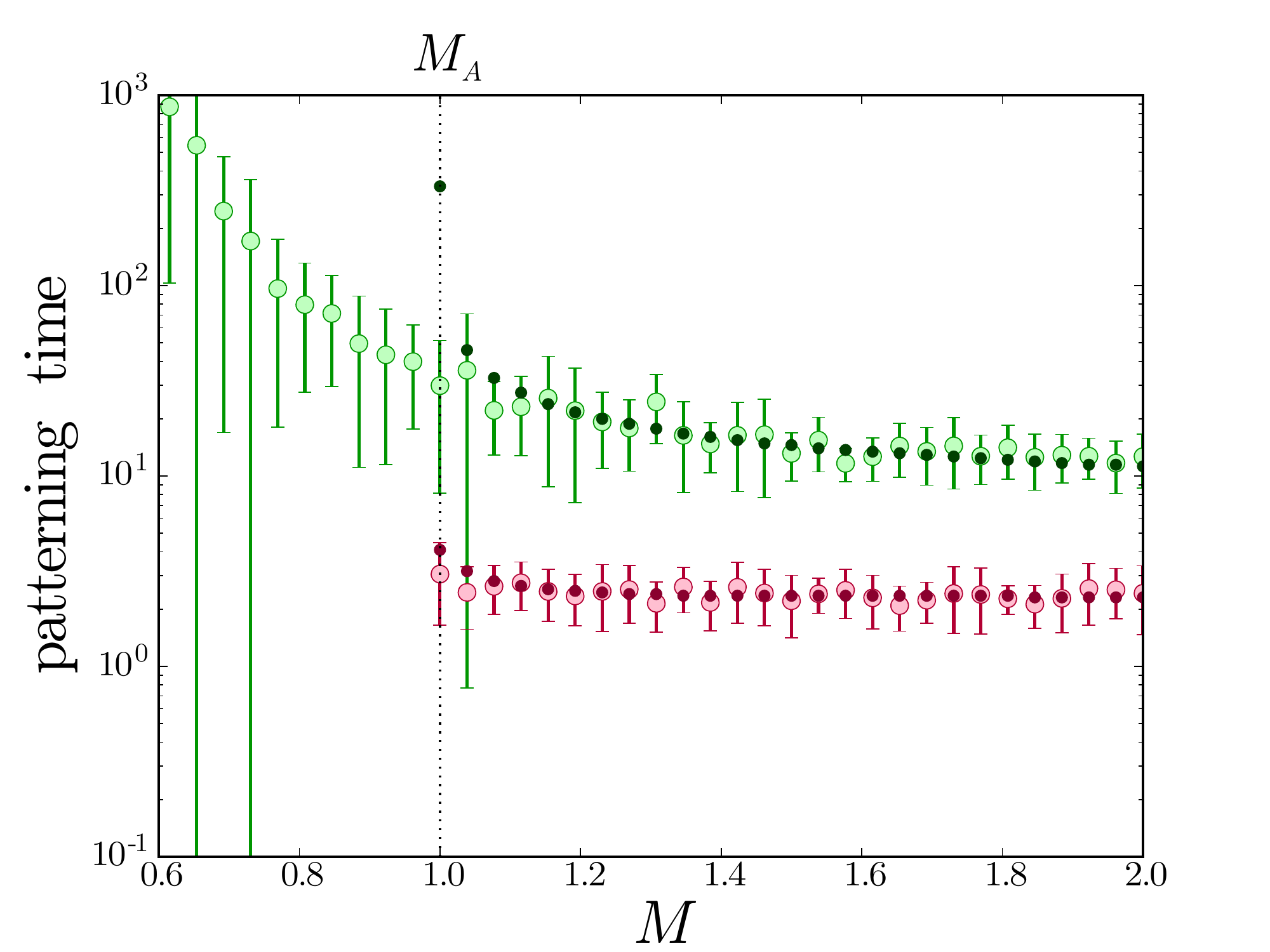}
\captionsetup{labelformat=empty}
\caption{{\bf Patterning circuit without feedback is not affected by intrinsic noise.} Patterning time for different levels of the signal for a case with intrinsic noise (big circles) and the deterministic case (small circles). The bistable switch (green circles) is compared with the non-feedback case (magenta circles). The patterning time is measured as the time necessary to observe a stationary response $x_A>0.9$. There is no patterning transient for the case without feedback for signal values below the $M_A$ (dotted line) since the initial state (phenotype B) is already the steady state phenotype. Each point corresponds to the average of 15 CLE realisations with standard deviation marked with error bars. Parameters are the same as in Fig. \ref{fig.transientprofilenoise}.}
\end{figure*}

\subsection*{S2 Fig}
\label{fig.signalboundaryOmegaOS}
\begin{figure*}[ht!]
\centering
\includegraphics[width=0.5\textwidth]{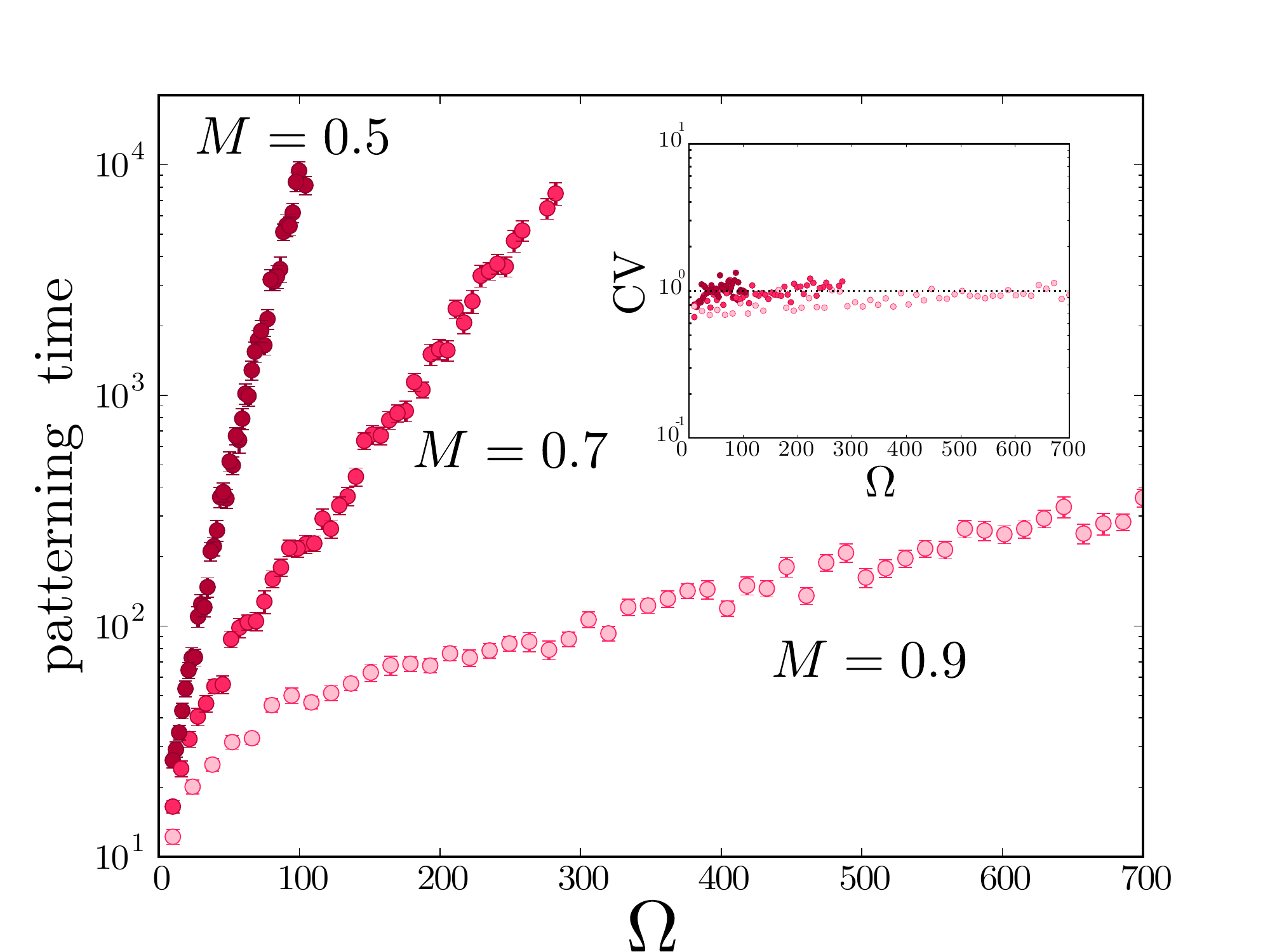}
\captionsetup{labelformat=empty}
\caption{
{\bf Mean patterning time as a function of the system size $\Omega$ inside the bistable zone.} Each point corresponds to the average mean first passage time of 100 CLE realisations. Inset) Coefficient of variation of the patterning times.  Error bars correspond with standard error of the mean. Parameters are the same as in Fig. \ref{fig.transientprofilenoise}.}
\end{figure*}

\clearpage
\subsection*{S3 Fig}
\label{fig.MAPsignal}
\begin{figure*}[ht!]
\centering
\includegraphics[width=0.7\textwidth]{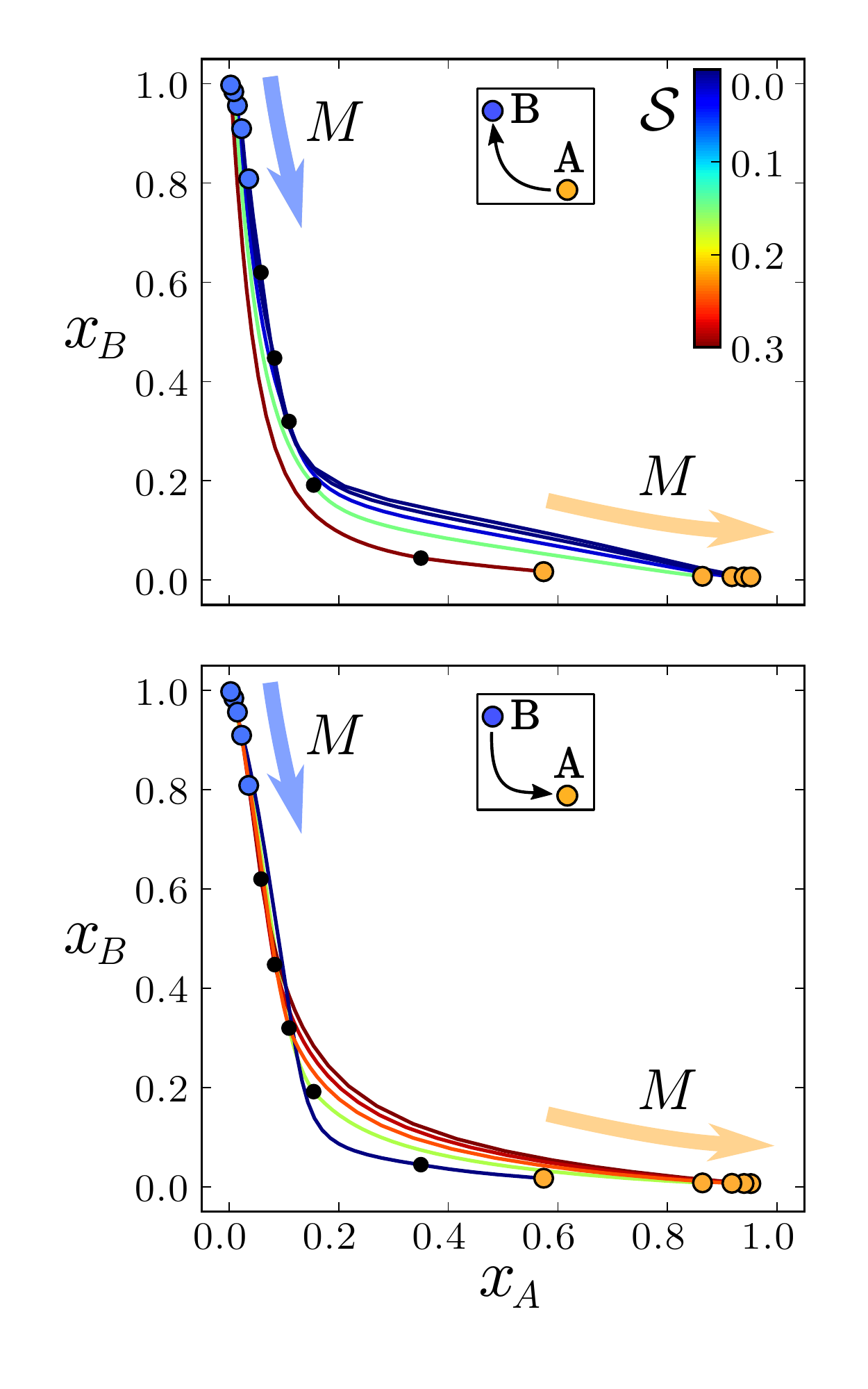}
\captionsetup{labelformat=empty}
\caption{{\bf Change of MAP along the bistable zone.} MAPs correspond to 5 different values of the signal $M$ for both transdifferentation processes $A\rightarrow B$ (top) and $B\rightarrow A$ (bottom). The value of the action (line color) changes with the signal. The position of the different steady expression states are marked with color circles for state A (orange) and B (blue) as well as the saddle points (black). The values of morphogen signal used are $M=$0.10,0.31,0.52,0.73,0.95.}
\end{figure*}

\clearpage
\subsection*{S4 Fig}
\label{MAPnuB}
\begin{figure*}[ht!]
\centering
\includegraphics[width=0.7\textwidth]{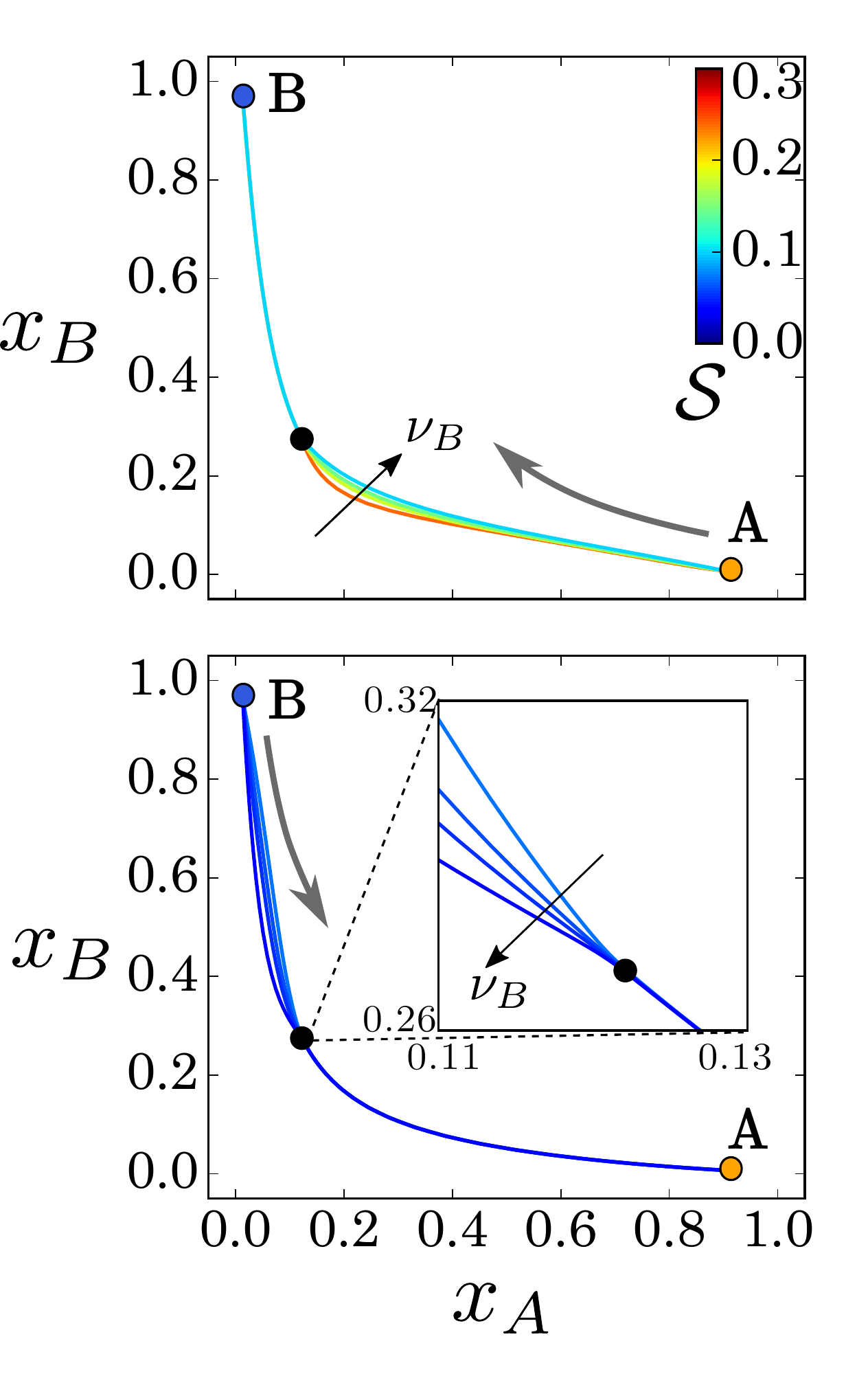}
\captionsetup{labelformat=empty}
\caption{
\textbf{Stochastic switching trajectory and rate change with the burst size $\nu_B$.} Change of MAP for different values of the relative burst size $\nu_B=1,3,5,10$ for both switching processes $A\rightarrow B$ (top) and $B\rightarrow A$ (bottom). The value of the action (line color) changes with $\nu_B$ . The position of the different steady expression states are marked with color circles for state A (orange) and B (blue) as well as the saddle points (black). Morphogen signal is  $M$=0.45. Parameters are those of Fig. \ref{fig.transientprofilenoise}.}
\end{figure*}

\clearpage
\subsection*{S5 Fig}
\label{fig.actionsnuA}
\begin{figure*}[ht!]
\centering
\includegraphics[width=0.7\textwidth]{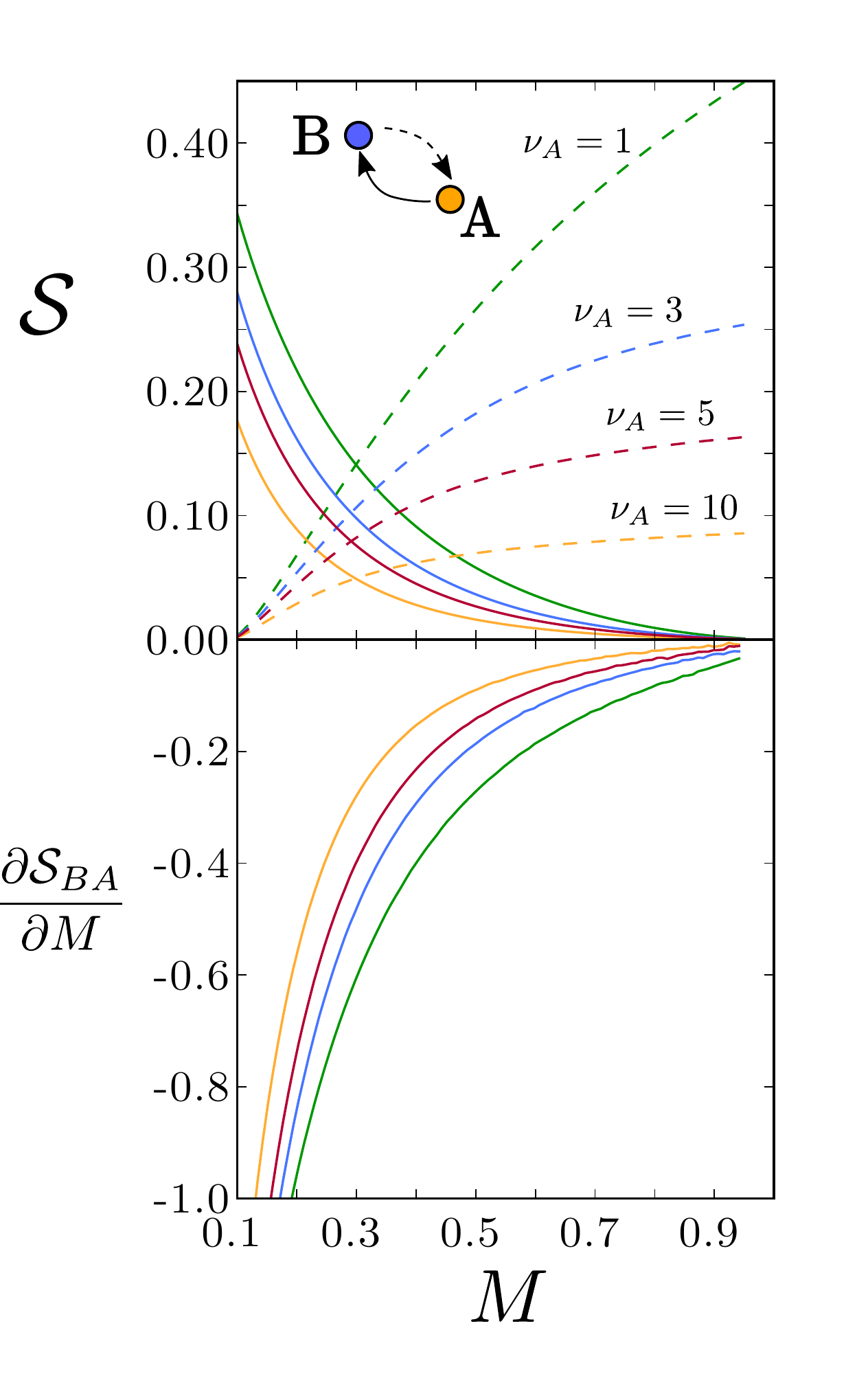}
\captionsetup{labelformat=empty}
\caption{\textbf{Burst size $\nu_A$ modifies the change of action along the tissue.} The action along the tissue is evaluated for different values of $\nu_A$ for both switching transitions: $B\rightarrow A$ (solid line) and $A\rightarrow B$ (dashed line), and its derivative over the morphogen concentration (bottom panel), vary as a function of the morphogen, revealing the time scale differences and directionality during the patterning process. Parameters are those of Fig. \ref{fig.transientprofilenoise}.}
\end{figure*}

\clearpage
\subsection*{S6 Fig}
\label{fig.velOmega}
\begin{figure*}[ht!]
\centering
\includegraphics[width=0.8\textwidth]{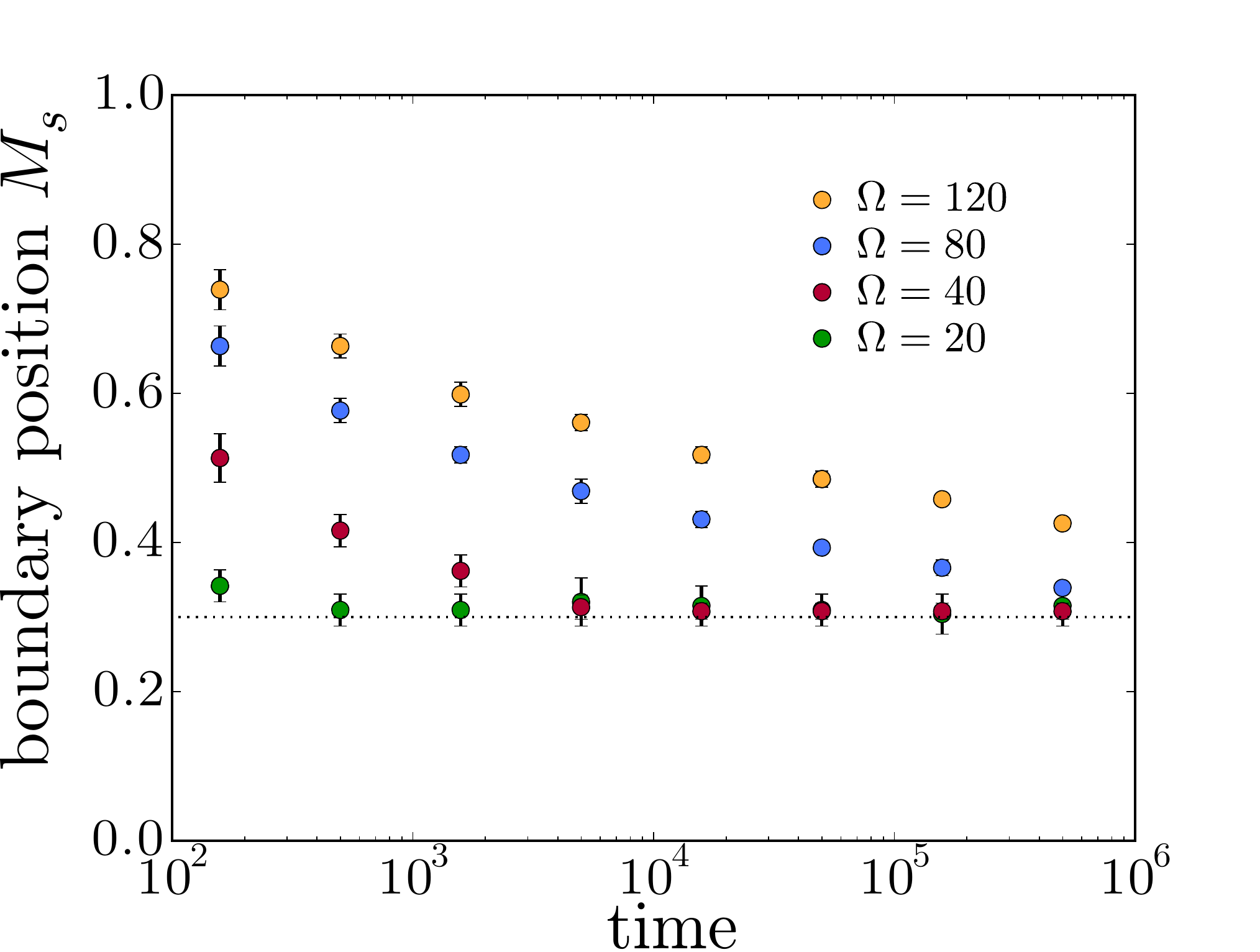}
\captionsetup{labelformat=empty}
\caption{
{\bf Position of the boundary towards the steady state depends on the typical number of proteins.} Position of the boundary is measured as the value of the signal for which $\langle x_A \rangle$=0.5 and error bars indicate the ranges $\langle x_A \rangle =[0.4,0.6]$. Each point is the average of 200 stochastic trajectories. Parameters are the same as in Fig. \ref{fig.transientprofilenoise}.}
\end{figure*}

\clearpage
\subsection*{S7 Fig}
\label{fig.velOmegaprofile}
\begin{figure*}[ht!]
\centering
\includegraphics[width=0.8\textwidth]{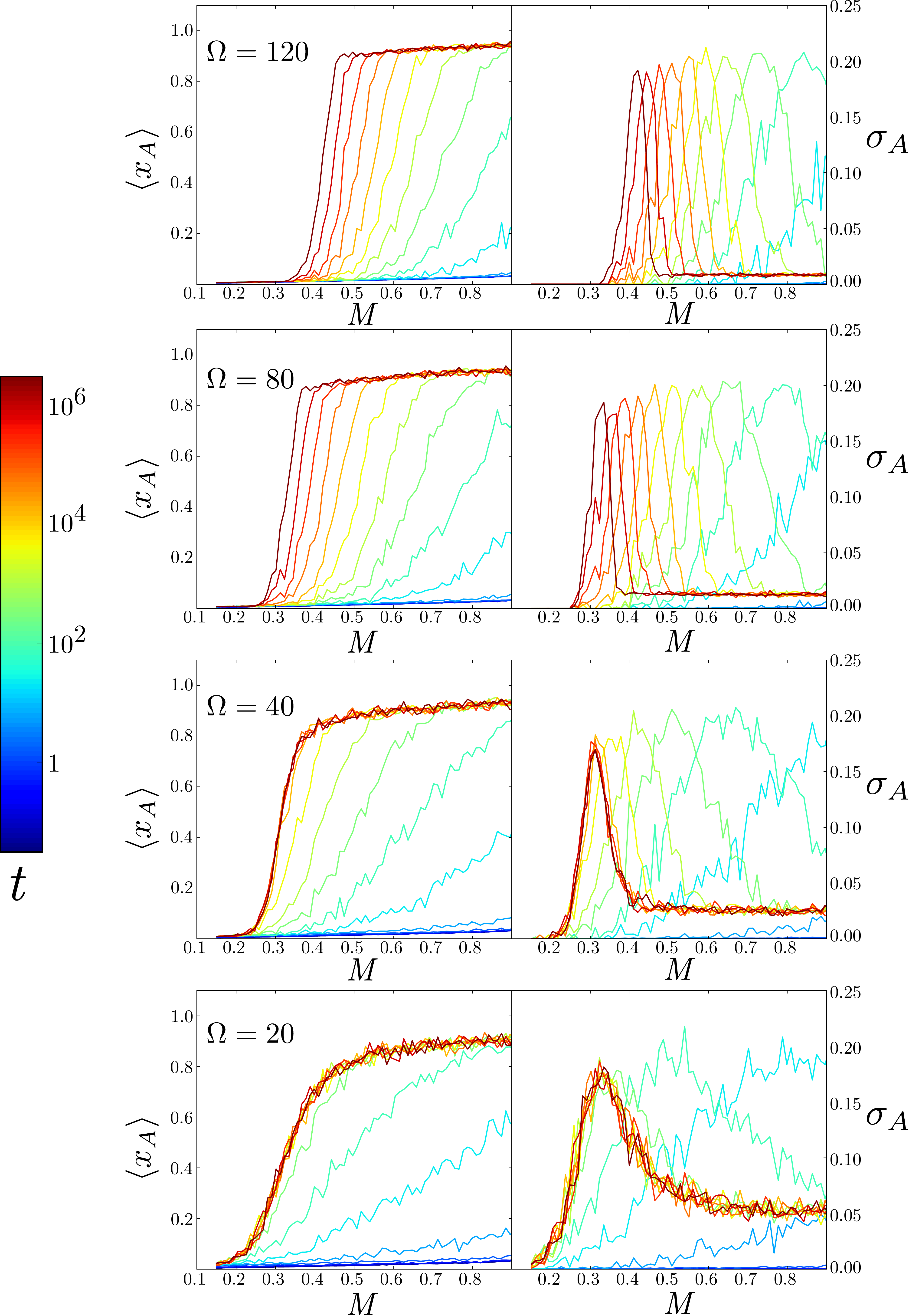}
\captionsetup{labelformat=empty}
\caption{{\bf  Travelling boundary velocity and precision depend on system size $\Omega$.} Mean and standard deviation in expression of the morphogen activated gene $A$ along the tissue at different time points for different values of $\Omega$. Results correspond to averaging of 500 trajectories with $\nu_A=\nu_B=1$; the rest of the parameters are the same as in Fig. \ref{fig.transientprofilenoise}. }
\end{figure*}

\end{document}